\documentclass[letterpaper,journal,final]{IEEEtran}

\usepackage{amssymb}
\usepackage{amsmath}
\usepackage{amsfonts}                                
\usepackage[latin1] {inputenc}
\usepackage{enumerate}
\usepackage{mathrsfs}
\usepackage{tabulary}
\usepackage{cite}
\usepackage{psfrag}
\usepackage{cite}
\usepackage{graphicx}          
\usepackage{color}      
\usepackage{epsfig} 
\usepackage{epstopdf}
\usepackage{times} 
\def\qedp{\hspace*{\fill}~{\tiny $\blacksquare$}}
\def\qed{\relax\ifmmode\hskip2em \Box\else\unskip\nobreak\hskip1em $\Box$\fi}

\newtheorem{theorem}{Theorem}
\newtheorem{itlemma}{Lemma}
\newtheorem{itdefinition}{Definition}
\newtheorem{itproposition}{Proposition}
\newtheorem{itresult}{Result}
\newtheorem{itremark}{Remark}
\newtheorem{itassumption}{Assumption}
\newtheorem{itcorollary}{Corollary}
\newtheorem{itexample}{Example}

\newenvironment{remark}{\begin{itremark}\rm}{\end{itremark}}
\newenvironment{assumption}{\begin{itassumption}\rm}{\end{itassumption}}
\newenvironment{lemma}{\begin{itlemma}\rm}{\end{itlemma}}

\title{\LARGE Dynamic Quantized Consensus of General Linear\\ Multi-agent Systems under Denial-of-Service Attacks} 

\author{
	Shuai Feng, Hideaki Ishii
	\thanks{Shuai Feng is with ENTEG, Faculty of Science and Engineering, University of Groningen, 9747AG Groningen, the Netherlands ({\tt\small s.feng@rug.nl}).}
		\thanks{ Hideaki Ishii is with the Department of Computer Science, Tokyo Institute of Technology, Yokohama 226-8502, Japan ({\tt\small ishii@c.titech.ac.jp}).}
	\thanks{		This work was supported in the part by the JST CREST Grant No.~JPMJCR15K3 and by JSPS under Grant-in-Aid for Scientific Research Grant No.~18H01460.}
}

\begin{document}
\maketitle
\begin{abstract}
In this paper, we study multi-agent consensus problems
under Denial-of-Service (DoS) attacks with data rate constraints.
We first consider the leaderless consensus problem and after that we briefly present the analysis of leader-follower consensus. 
The dynamics of the agents take general forms modeled
as homogeneous linear time-invariant systems.
In our analysis, we derive lower bounds on the data rate
for the multi-agent systems to achieve leaderless and leader-follower consensus
in the presence of DoS attacks, under which the issue of overflow of quantizer is prevented. The main contribution
of the paper is the characterization of the trade-off
between the tolerable DoS attack levels for leaderless and leader-follower consensus and the required
data rates for the quantizers during the communication attempts among the agents.
To mitigate the influence of DoS attacks, we employ dynamic quantization
with zooming-in and zooming-out capabilities for avoiding quantizer saturation.	
\end{abstract}


\section{Introduction}
In the last two decades, the control of multi-agent systems has attracted substantial attention due to the progress of technologies in communication and computation areas. Some of the key applications can be found in formation control, control of large-scale systems and distributed sensor networks \cite{FB-LNS}. In particular, nowadays a closed-loop control system integrates sensors, computers and communication devices, which complies with the concept of cyber-physical systems (CPSs). While the industry notably benefits from the technology bloom in CPSs, a challenging situation also emerges along with the benefits due to malicious cyber attacks on CPSs \cite{cheng2017guest, teixeira2015secure, Lu, wang2020privacy}, in the form of, e.g., deceptive attacks and Denial-of-Service (DoS).

This paper specifically deals with DoS attacks, which induce packet drops maliciously and hence corrupt the availability of data. The communication failures induced by DoS can exhibit a temporal profile quite different from those caused by genuine packet losses due to 
network congestion; particularly packet dropouts resulting from malicious DoS need not follow a given class of probability distributions \cite{sastry}, and therefore the analysis techniques relying on probabilistic arguments may not be applicable. This poses new challenges in theoretical analysis and controller design.

In this paper, our focus is on the effects of DoS attacks on multi-agent
systems. Recently, systems under such attacks have been studied from
a control-theoretic viewpoint \cite{de2015input,  8049303, feng2020tac, li2017sinr, automatica2017, senejohnny2017jamming, 8353464, feng2019secure, feng2016distributed1, xu2019distributed, Deng}. 
In \cite{de2015input}, a framework is introduced where DoS attacks are characterized by their levels of \emph{frequency} and \emph{duration}. 
There, they derived an explicit characterization of DoS frequency and duration under which stability can be preserved
through state-feedback control. For multi-agent systems under DoS, there are some recent results for consensus problems with infinite data-rate communication. For example, the paper \cite{feng2019secure} presents theoretical as well as comprehensive simulation studies for continuous-time system consensus under DoS attacks with the utilization of event-triggered control.

Wireless communication appeals to industry
due to the advantages such as transmission over
long distances and lower costs for large-scale implementation.
However, the transmitted signals are subject to analog-digital conversion and hence
quantization. Real-time data exchanged within networked control systems may suffer from data rate constraints and hence the quantization effects on control/measurement signals need to be taken into account at the design stage. 
\color{black}
Static and dynamic quantizations have been proposed for various control problems. Centralized control systems under quantized communication have been extensively studied in the last two decades, for example by the seminal papers \cite{1310461, nair2004stabilizability} and the book \cite{ishii2002limited}.
The results in such works show that insufficient bit rate in communication channel influences the stability of a networked control system. The paper \cite{feng2020tac} extended these results to the case with DoS attacks by utilizing zooming-in and out dynamic quantization for centralized systems. 
In this paper, we address issues arising from constraints on data rate that can occur in multi-agent systems.

In addition to centralized systems, quantized consensus problems of multi-agent systems have been broadly studied in the last decade\cite{wang2019event, carli2010gossip, kashyap2007quantized, you2011network, qiu2015quantized} and some of them take data rate constraints into considerations. The related problem
of quantized resilient consensus is studied in \cite{wang2019event} where some agents are malicious and may prevent consensus to take place. 
Indeed, the issue of data rate in networked control of multi-agent systems will become relevant
especially if the agents try to reach consensus on
multiple variables and the volume of data communicated among
agents is large. Reducing the data size for each variable
is one way to make the system robust to changes in
available data rate, e.g., in wireless communication.
In \cite{you2011network}, the authors consider the zooming-in only quantized controller with finite data rate. However, such a controller may not be feasible in the context of DoS since the quantizer would have overflow problems under DoS, due to state divergence. This is one of the central problems caused by considering data rate/quantization when investigating a resilient control problem. In order to mitigate the influence of DoS attacks and to ensure that a channel with finite data rate is still feasible, in this paper, we design a quantized controller with both \emph{zooming-in} and \emph{zooming-out} capabilities. We will show that in the absence of DoS attacks, our result in the part of leaderless consensus is consistent with the one in \cite{you2011network}. 


In light of the existing literature mentioned above and the comparisons, we summarize the contributions of this paper. Our work addresses the joint effects of DoS attacks and data rate constraints for both the leaderless and leader-follower consensus problems:
\begin{itemize}
	\item We explicitly demonstrate the trade-offs between \emph{the resilience against
		DoS} and \emph{the necessary data rate in communication}. That is, we find a data-rate dependent bound of DoS \emph{frequency} and \emph{duration} under which consensus can be achieved. Such trade-offs can provide useful information for the allocation of communication resources, e.g., to ensure that the multi-agent systems can realize the global objective of consensus under DoS, how much data rate must be allocated to the channel; and to improve the resilience, how much additional data rate must be ensured and so on. 
	\item  We develop the zooming-in and zooming-out dynamic quantization for the case of multi-agent systems. Specifically, we provide the sufficient number of quantization levels and the resulting bit rate, and particularly introduce the computation of zooming-out factor counteracting packet dropouts. They together ensure that the encoding-decoding systems are free of overflow under DoS induced packet losses. 
\end{itemize}



We now make more specific comparisons with existing works. 
As mentioned above, the paper \cite{feng2019secure} considers consensus
under DoS attacks with infinite data rates for communication.
There, the sufficient 
condition on DoS attacks for reaching consensus mainly depends
on the properties of the multi-agent systems (e.g., the system matrix $A$ and consensus rate during DoS-free periods).
In contrast, our paper incorporates
the constraints on data rate and develop
encoding and decoding systems functioning even in
the presence of DoS. In this case, the system resilience
also depends on data rate.


The computation of zooming-out factor for
multi-agent systems is one of the key technical
challenges in this paper. This issue arises due to the lack of 
``global state information'' to the agents
(when the network forms a non-complete graph).
For the centralized system case in \cite{feng2020tac}, such information 
is in fact useful in the zooming-in and out dynamic
quantization applied there. 
In the case of multi-agent systems, the encoding-decoding
system of a single agent cannot have the information about
its neighbors' states and also control inputs under DoS (since control inputs
of its neighbors also depend on their own neighbors). This
lack of information induces considerable technical difficulties
for tracking the states of neighbors, and hence for the design
of the zooming-out factor.


\color{black}

This paper is organized as follows. In Section II, we introduce the framework consisting of multi-agent systems and the class of DoS attacks. Section III presents the results of leaderless consensus, which includes the controller architecture with the zooming-in and zooming-out dynamic quantization mechanism and sufficient conditions for data rates and DoS bounds under which consensus can be achieved. Section IV briefly presents an extension of the results to the leader-follower consensus problem.  
A numerical example is presented in Section V, and finally Section VI ends the paper with conclusions and possible future research directions. 
This paper mainly focuses on the case of leaderless consensus, which provides the theoretical foundations for the part of leader-follower consensus.
Preliminary results for quantized leaderless and leader-follower consensus under DoS can be found in our conference papers \cite{feng2020ifac} and \cite{feng2020cdc}, respectively. Compared with them, this paper provides full proofs of the results, more discussions and comparisons.

\textbf{Notation.} We denote by  $\mathbb R$ the set of reals. Given $b \in \mathbb R$, $\mathbb R_{\geq b}$ and $\mathbb R_{>b}$ denote the sets of reals no smaller than $b$ and reals greater than $b$, respectively; $\mathbb R_{\le b}$ and $\mathbb R_{<b}$ represent the sets of reals no larger than $b$ and reals smaller than $b$, respectively; $\mathbb Z$ denotes the set of integers. For any $c \in \mathbb Z$, we denote $\mathbb Z_{\ge c} := \{c,c+ 1,\cdots\}$. Let $\lfloor v \rfloor$ be the floor function such that $\lfloor v \rfloor= \max\{o\in \mathbb{Z}|o\le v\}$. 
Given a vector $y$ and a matrix $\Gamma$, let $\|y\|$ and $\|y\|_\infty$ denote the $2 $- and $\infty$- norms of vector $y$, respectively,  
and $\|\Gamma\|$ and $\|\Gamma\|_\infty$ represent the corresponding induced norms of matrix $\Gamma$. $\rho(\Gamma)$ denotes the spectral radius of $\Gamma$. Given an interval $\mathcal{I}$, $|\mathcal{I}|$ denotes its length. The Kronecker product is denoted by $\otimes$. Let $\mathbf 0$ and $\mathbf 1$ denote the column vectors with compatible dimensions, having all $0$ and $1$ elements, respectively.

\section{Framework: multi-agent systems and DoS}
\subsection{Communication graph}
We let graph $\mathcal{G} = (\mathcal{V},\mathcal{E})$ denote the communication topology between agents, where $\mathcal{V}=\{1, 2, \cdots, N \}$ denotes the set of agents and $\mathcal{E} \subseteq \mathcal{V} \times \mathcal{V}$ denotes the set of edges. Let $\mathcal N_i$ denote the set of the neighbors of agent $i$, where $i=1, 2, \cdots, N$. In this paper, we assume that the graph $\mathcal{G}$ is undirected and connected, i.e. if $j \in \mathcal N_i$, then $i \in \mathcal N_j$. Let $A _ {\mathcal G}= [a_{ij}] \in \mathbb{R}^{N\times N} $ denote the adjacency matrix of the graph $\mathcal {G}$, where $a_{ij} > 0$ if and only if $j \in \mathcal N_i$ and $a_{ii}=0$. Define the Laplacian matrix $L_ \mathcal{G} = [l_{ij}]  \in \mathbb{R}^{N\times N} $, in which $l_{ii} = \sum_{j = 1 }^{N} a_{ij}$ and $l_{ij} = - a_{ij} $ if $i \ne j $. Let $\lambda_i$ ($i=1, 2, \cdots, N$) denote the eigenvalues of $L_ \mathcal{G}$ and in particular we have $\lambda_1 = 0$ due to the graph being connected.  

\subsection{System description}

The agents with interacting over the network $\mathcal{G}$ are expressed as homogeneous general linear multi-agent systems. For each $i=1,2, \cdots, N$, agent $i$ is given as a sampled-data system with sampling period $\Delta \in \mathbb{R}_{>0} $ in the form of
\begin{align}\label{system}
x_i(k\Delta) = A x_i((k-1)\Delta) + B u_i((k-1)\Delta)
\end{align}
where $k\in \mathbb{Z}_{\ge 1}$, $A\in \mathbb R^{n \times n}$ and $B\in \mathbb R^{n \times w}$. It is assumed that $(A,B)$ is stabilizable. 
$x_i(k \Delta)\in \mathbb{R}^{n}$ denotes the state of agent $i$ with $x_i(0) \in  \mathbb{R}^{n}$ as the initial condition. We assume that an upper bound is known, i.e. $\| x_i(0)\|_\infty \le C_{x_0} \in \mathbb{R}_{>0}$. 
Note that $C_{x_0}$ can be an arbitrarily large real as long as it satisfies this bound. This is for preventing the overflow of state quantization for the initial condition.
Let $u_i((k-1)\Delta) \in \mathbb{R}^{w}$ denote its control input, whose computation will be given later. 

We assume that the communication channel among the agents is bandwidth limited and subject to DoS, 
where transmission attempts take place periodically at time $k\Delta$ with $k\in \mathbb{Z}_{\ge 1}$. Moreover, we assume that the transmission is acknowledgment based and free of delay. This implies that the decoders send acknowledgments to the encoders immediately when they receive encoded signals successfully. If some acknowledgments are not received by the encoders, it implies that due to the presence of DoS, the decoders do not receive any data, and hence they do not send acknowledgments. 

Agent $i = 1, 2, \cdots, N$ can only exchange information with its neighbor agents $j \in \mathcal N_i$. Due to the constraints of network bandwidth, signals are encoded with a limited number of bits. In the presence of DoS, transmission attempts may fail. For the ease of notation, we let $s_r$ represent the instants of successful transmissions. Note that $s_0 \in \mathbb{R}_{\ge \Delta}$ is the instant when the first successful transmission occurs. Also, we let $s_{-1}$ denote the time instant $0$. 

\textbf{Uniform quantizer.}
The limitation of bandwidth implies that transmitted signals are subject to quantization. Let $\chi \in \mathbb{R}$ be the original scalar value before quantization and $q_R(\cdot)$ be the quantization function for scalar
input values as
\begin{align}\setlength{\arraycolsep}{4pt}  \label{quantizer}
q_R (\chi) = 
\left\{
\begin{array}{lll}
0 & -\sigma < \chi < \sigma & \\
2z \sigma & (2z-1)\sigma \le \chi < (2z+1)\sigma  \\
2R\sigma & \chi \ge  (2R+1) \sigma&         \\
-q_R (-\chi) & \chi \le - \sigma & 
\end{array}
\right.
\end{align}
where $R\in \mathbb{Z}_{>0}$ is to be designed and $z =1, 2, ..., R$, and $\sigma \in \mathbb{R}_{>0}$. If the quantizer is unsaturated such that $\chi \le (2R+1)\sigma $, then the error induced by quantization satisfies 
\begin{align}\label{quantizer error}
|\chi - q_R(\chi)| \le \sigma, \,\, \text{if}\,\, |\chi| \le (2R+1)\sigma.
\end{align}
Observe that the quantizer has $2R+1$ levels and is determined by two parameters $\sigma$ and $R$, which determine the density and quantization range of the quantizer, respectively. Moreover, we define the vector version of the quantization function as $Q_R(\beta) = [\,q_R(\beta_1)\,\,q_R(\beta_2)\,\, \cdots \,\, q_R(\beta_f) \, ]^T \in \mathbb R ^f$, where $\beta = [\beta_1\,\, \beta_ 2\,\, \cdots \beta_f]^ T \in \mathbb R ^f$ with $f \in \mathbb Z_{\ge 1} $.

To design safe control systems, we must make assumptions
regarding the DoS attacks that we expect the systems with sufficient safety margins. 
If the attackers are more capable than the assumed
attack level, it would clearly be hard to guarantee
consensus. This will however be true for any model. In the next subsection, we introduce a deterministic model characterizing DoS attacks. This allows us to consider the worst-case attacks, without assuming any probability distributions for launching attacks as in the random packet loss model commonly studied in the networked control literature.

\subsection{Time-constrained DoS}
In this paper, we refer to DoS as the event for which all the encoded signals cannot be received by the decoders and it affects all the agents.  
We consider a general DoS model
that describes the attacker's action by the frequency of DoS attacks and their duration. Let 
$\{h_q\}_{q \in \mathbb Z_0}$ with $h_0 \geq \Delta$ denote the sequence 
of DoS \emph{off/on} transitions, that is,
the time instants at which DoS exhibits 
a transition from zero (transmissions are successful) to one 
(transmissions are not successful).
Hence,
$
H_q :=\{h_q\} \cup [h_q,h_q+\tau_q[  
$
represents the $q$-th DoS time-interval, of a length $\tau_q \in \mathbb R_{\geq 0}$,
over which the network is in DoS status. If $\tau_q=0$, then
$H_q$ takes the form of a single pulse at $h_q$.  
Given $\tau,t \in \mathbb R_{\geq0}$ with $t\geq\tau$, 
let $n(\tau,t)$
denote the number of DoS \emph{off/on} transitions
over $[\tau,t]$, and let 
$
\Xi(\tau,t) := \bigcup_{q \in \mathbb Z_0} H_q  \, \cap  \, [\tau,t] 
$
be the subset of $[\tau,t]$ where the network is in DoS status. 

\begin{assumption}
	(\emph{DoS frequency}). 
	There exist constants 
	$\eta \in \mathbb R_{\geq 0}$ and 
	$\tau_D \in \mathbb R_{> 0}$ such that
	\begin{align} \label{ass:DoS_slow_frequency} 
	n(\tau,t)  \, \leq \,  \eta + \frac{t-\tau}{\tau_D}
	\end{align}
	for all  $\tau,t \in \mathbb R_{>0}$ with $t\geq\tau$.
	\qedp
\end{assumption}

\begin{assumption} 
	(\emph{DoS duration}). 
	There exist constants $\kappa \in \mathbb R_{\geq 0}$ and $T  \in \mathbb R_{>1}$ such that
	\begin{align} \label{ass:DoS_slow_duration}
	|\Xi(\tau,t)|  \, \leq \,  \kappa + \frac{t-\tau}{T}
	\end{align}
	for all  $\tau,t \in \mathbb R_{>0}$ with $t\geq\tau$. 
	\qedp
\end{assumption}

\begin{remark}
	Assumptions 1 and 2
	do only constrain a given DoS signal in terms of its \emph{average} frequency and duration.
	Following \cite{hespanha1999stability}, 
	$\tau_D$ can be defined as the average dwell-time between 
	consecutive DoS off/on transitions, while $\eta$ is the chattering bound.
	Assumption 2 expresses a similar 
	requirement with respect to the duration of DoS. 
	It expresses the property that, on the average,
	the total duration over which communication is 
	interrupted does not exceed a certain \emph{fraction} of time,
	as specified by $1/T$.
	Like $\eta$, the constant $\kappa$ plays the role
	of a regularization term. It is needed because
	during a DoS interval, one has $|\Xi(h_q,h_q+\tau_q)| = \tau_q >  \tau_q /T$.
	Thus $\kappa$ serves to make (\ref{ass:DoS_slow_duration}) consistent. 
	Conditions $\tau_D>0$ and $T>1$ imply that DoS cannot occur at an infinitely
	fast rate or be always active. \qedp
\end{remark}

The next lemmas relate DoS parameters and the number of unsuccessful and successful transmissions, respectively.
\begin{lemma}\label{Lemma Q}
	Consider a periodic transmission with sampling interval $\Delta$
	along with DoS attacks under Assumptions 1 and 2. If 
	$1/T+ \Delta /\tau_D <1$, then $m_r$, representing the number of unsuccessful transmissions between $s_{r-1}$ and $s_r$ with $r= 0, 1, \cdots$,
	satisfies  
	\begin{align}
	m_r &= (s_{r} - s_{r-1})/\Delta -1 \nonumber\\
	&\le M = \left\lfloor \frac{(\kappa + \eta \Delta) \left(1- 1/T  - \Delta / \tau _D  \right)^{-1}  }{\Delta} \right \rfloor \in \mathbb  Z _{\ge 0}.
	\end{align}
\end{lemma}

\emph{Proof}.
This lemma can be easily derived from Lemma 1 in \cite{automatica2017} and we refer the readers to the full proof there. 
\qedp 

For the ease of notation, we let $m$ represent $m_r$ in the subsequent sections.  

\begin{lemma} \label{Lemma T}
	Consider the DoS attacks characterized by Assumptions 1 and 2 and the network sampling period $\Delta$. If 
	$1/T+ \Delta /\tau_D <1$, then $T_S(\Delta, k\Delta)$, denoting the number 
	of successful transmissions within
	the interval $[\Delta, k\Delta]$,
	satisfies
	\begin{align}\label{Ts}
	T_S(\Delta, k \Delta)
	\ge  \left(1- 1/T - \Delta/\tau_D \right) k   -  (\kappa+\eta\Delta)/\Delta.
	\end{align} 
\end{lemma}

\emph{Proof.} This lemma can be easily derived from Lemma 3 in \cite{feng2020tac} and we refer the readers to that paper. \qedp

\begin{remark}
	If the network is free of DoS attacks ($T= \tau_D = \infty$ and $\kappa = \eta =0$), then $m=M=0$ and $T_S(\Delta, k \Delta) = k$, i.e., there is no failure in transmissions between $s_{r-1}$ and $s_r$ for every $r$, and every transmission attempt will be successful, respectively. Therefore, they reduce to nominal standard periodic transmissions. \qedp    
\end{remark}

\section{Leaderless quantized consensus under DoS}

The objective of this section is to design a quantized controller, possibly dynamic, in such a way that a finite-level quantizer is not overflowed and the multi-agent system (\ref{system}) can tolerate as many DoS attacks as possible for reaching consensus.
Specifically, we introduce 
the average of the states
$
\overline{x}(k\Delta) = \frac{1}{N}\sum_{i=1}^N x_i(k\Delta)   \in \mathbb{R}^{n}
$
and consensus among the agents is defined by 
\begin{align}\label{control objective}
\lim_{k  \to \infty } \| x_i(k\Delta) - \overline  x(k\Delta)  \| _ \infty = 0, \,\,\, i=1,2,\cdots, N.
\end{align}
For the ease of illustration, in the remainder of the paper we simply let $k$ represent $k\Delta$, e.g., $x_i(k)$ represents $x_i(k\Delta)$. 
We are interested in some $A$ having at least one eigenvalue
on or outside the unit circle. Otherwise, the multi-agent system in (\ref{system})
can achieve state consensus by setting $u_i(k) = 0$
for all $k$.

\subsection{Control architecture for leaderless consensus}
For each agent $i$, the control input $u_i(k)$ is expressed as a function of the relative states available locally at time $k$. Specifically, it is given by
\begin{align}\label{controller 0}
u_i(k) = K \sum_{j=1} ^{N} a_{ij} (\hat x_j ^ i (k) - \hat x_i ^ i (k)), \,\,k=0, 1, \cdots,
\end{align}
where $\hat x _ j ^ i(k) \in \mathbb{R}^n$ denotes the estimation of the state of agent $j$ by agent $i$, whose computation will be given later. 
We assume that there exists a feedback gain $K \in \mathbb R ^{w \times n}$ such that the spectral radius of
\begin{align}\label{J(1)}
J(1)= \text{diag}(A- \lambda_2 BK,  \cdots, A- \lambda_N BK)
\end{align}
satisfies $\rho(J(1)) < 1$. This is a standard
condition for consensus when no 
DoS is present \cite{li2009consensus, you2011network}.

In (\ref{controller 0}), 
the estimate of the state of agent $j$ by agent $i$ equals the one estimated by agent $l$
such that $\hat x _ j ^ i(k) = \hat x _ j ^ l (k) = \hat x_j ^ j (k)$ with $i,  l \in \mathcal N_j$. Thus, we omit the superscripts and let
\begin{align}\label{controller}
u_i(k) = K \sum_{j=1} ^{N} a_{ij} (\hat x_j  (k) - \hat x_i (k)), k=0, 1, \cdots.
\end{align}
Agent $i$ estimates the states of its neighbors based on the information available from communication.
Also, to stay consistent with the neighbors, it will compute
the estimate of its own. 
These estimated states will be computed at each time $k=1, 2, \cdots$ as
\begin{align}\setlength{\arraycolsep}{3pt}  \label{eq estimator}
\hat x_j(k)
=
\left\{
\begin{array}{ll}
A \hat x_j(k-1) + \theta(k-1) \hat Q_j(k)  & \text{if $k \notin H_q$ } \\
A \hat x_j(k-1)               & \text{if $k \in  H_q$} 
\end{array}
\right.
\end{align}
where $j\in\{i\}\cup\mathcal{N}_i$ and the initial estimates will be set as $\hat x_j (0)= \mathbf 0$.
The scaling parameter $\theta(k) \in \mathbb{R}_{>0}$ in (\ref{eq estimator}) is updated as
\begin{align}\label{eq h}
\theta(k) = 
\left\{
\begin{array}{ll}
\gamma_1 \theta(k-1) & \quad \text{if $k  \notin H_q $} \\
\gamma_2 \theta(k-1) & \quad \text{if $k  \in H_q $  }
\end{array}
\right.  \,\,\, k=1, 2, \cdots
\end{align}
with $\theta(0)  = \theta_0 \in \mathbb{R}_{>0}$, where $0<\gamma _ 1<1$ and $\gamma _2 > 0$. 
Moreover, the scaling parameter $\theta(k)$ is used in the quantization $\hat Q_j(k)$ given by 
\begin{align}\label{Q_i}
\hat Q_j(k) = Q_R \left(\frac{x_j(k) -  A \hat x_j(k-1)}{\theta(k-1)}  \right), \,\, k=1, 2, \cdots
\end{align}
for preventing quantizer overflow. 
By adjusting the size of $\theta(k)$ dynamically, the state will be kept within the bounded quantization range without saturation, i.e., $\frac{x_j(k) -  A \hat x_j(k-1)}{\theta(k-1)}$ in $Q_R(\cdot)$ is upper bounded by some certain values. 
The parameters $\gamma_1$ and $\gamma_2$ in (\ref{eq h}) are for zooming in and out such that the quantization scaling parameter $\theta(k)$ changes dynamically to mitigate the influence of DoS.  
Under DoS attacks, the states of the multi-agent systems may diverge. Therefore, the quantizers must zoom out and increase their ranges so that the states can be measured properly. If the transmissions succeed, the quantizers zoom in and $\theta(k)$ decreases by using $\gamma_1$.   
The design of $\gamma_1$, $\gamma_2$ and $\theta_0$ will be specified later.

Observe that the controller state is updated locally at each agent by checking the presence of DoS attacks over time. It is clear that each agent has access to the knowledge
of DoS attacks in real time from not receiving data from the
neighbors at the scheduled periodic transmission instants. One sees that the estimator (\ref{eq estimator}) switches the estimation strategy adaptively to the information if $\hat Q_j(k)$ is available to the controller ($k  \notin H_q $) or not ($k  \in H_q $). In particular, if $\hat Q_j(k)$ is lost, then set $\hat Q_j(k)=0$. The ``to zero" strategy is commonly used in networked control problems with information loss. Note that the calculation of $\hat Q_j(k)$ (at the encoder) is dependent on $\theta(k-1)$ that needs the past information of $k-1  \notin H_q $ or $k-1  \in H_q$, instead of the corresponding information for $k$.

The overall estimation and update processes are summarized as follows. 
The state $x_j(k)$ is quantized into $\hat Q_j(k)$ as in (\ref{Q_i}) by the encoder and agent $j$ attempts to send it to the decoders of its neighbors. If the transmission attempt succeeds and $\hat Q_j(k)$ is received, the decoders estimate $x_j(k)$ by the first equation in (\ref{eq estimator}) and the scaling parameter $\theta(k)$ in the encoders and decoders zooms in by the first equation in (\ref{eq h}). If the transmission attempt fails, the information of $x_j(k)$ cannot be acquired by the decoders since $\hat Q_j(k)$ is corrupted by DoS. Then, the decoders estimate $x_j(k)$ by the second equation in (\ref{eq estimator}) and the scaling parameter $\theta(k)$ 
in the encoders and decoders zooms out as in the second equation in (\ref{eq h}).

Note that in the control input (\ref{controller}), we use $\hat x_i(k)$ to compute $u_i(k)$ instead of $x_i (k)$. Due to space limitation, we omit the details of the rationales and refer the readers to the discussion regarding (52) in \cite{you2011network} and the references therein.

Let $\hat x(k) = [\hat x_1 ^ T (k) \,\, \hat x_2 ^ T (k) \cdots \hat x_N ^ T (k)]^T \in \mathbb{R}^{nN}$ and $Q(k) = [\hat Q_1 ^ T (k) \,\,\hat Q_2 ^ T (k) \cdots \hat Q_N ^ T (k)]^T  \in \mathbb{R}^{nN}$. One can obtain the compact form of (\ref{eq estimator}) as
\begin{eqnarray}\setlength{\arraycolsep}{1pt}  \label{eq estimator all}
\hat x(k)  
=
\left\{
\begin{array}{ll}
(I_N \otimes A) \hat x(k-1) + \theta(k-1)Q(k) & \text{if $k \notin H_q$} \\
(I_N \otimes A) \hat x(k-1)        & \text{if $k \in H_q$} 
\end{array}
\right.
\end{eqnarray}
for $k=1, 2, \cdots$.
Let $e_i(k) = x_i(k) - \hat x_i(k) \in \mathbb{R}^{n} $ denote the estimation error and let $e(k) = [e_1 ^T(k) \,\, e_2 ^T(k)\,\, \cdots \,\, e_N ^T(k)]^T \in \mathbb{R}^{nN}$ and $x(k) = [x_1 ^ T (k) \,\, x_2 ^ T (k) \cdots x_N ^ T (k)]^T \in \mathbb{R}^{nN}$. Then one obtains the compact form of the dynamics of the agents 
\begin{align}\label{eq process all}
x(k) = G x(k-1)  + Le (k-1)
\end{align}
where
\begin{align}\label{G}
G = I_N \otimes A - L_ \mathcal{G}\otimes B K,\,\,\, L= L_ \mathcal{G} \otimes BK.
\end{align}

Recall the average of the states $\overline x(k)$ before (\ref{control objective}). The discrepancy between the state of agent $i$ and $\overline{x}$ is denoted by $\delta_i(k) = x_i(k) - \overline x(k) \in \mathbb{R}^{n}$.
By defining $\delta(k) = [\delta_1 ^T (k) \,\, \delta_2 ^T (k)\,\, \cdots \,\, \delta_N ^T (k)]^T \in \mathbb{R}^{nN}$, one has $x(k) = \delta (k) + I_N \otimes \overline x(k)$. By applying it to (\ref{eq process all}), one obtains
\begin{align}\label{delta all}
\delta(k) =   G \delta(k-1) +L e (k-1).
\end{align}

It is clear that if $\|\delta (k)\|_\infty \to 0$ as $k \to \infty$, consensus of the multi-agent system (\ref{system}) is achieved as in (\ref{control objective}).
If $\|e(k)\| =0$ or is upper bounded by a certain value \cite{you2011network} for all $k$, it is obvious that consensus can be achieved. 
Under DoS attacks, however, $e(k)$ may diverge and consequently consensus among the agents may not be achieved.

\subsection{Dynamics of the multi-agent systems}
In this subsection, we present the dynamics of the multi-agent system under quantization, in terms of $e(k)$ with $e(k-1)$ and $\delta(k-1)$ for the two cases, i.e., in the absence and presence of DoS attacks.

\emph{If the transmission succeeds} such that $k \notin H_q$ for $k= 1, 2, \cdots$, then according to (\ref{eq estimator all}), one has
\begin{align}\label{e no DoS 1}
e(k) 
=&\, x(k) - \hat x(k) \nonumber\\
=&\, x(k) - (I_N \otimes A) \hat x(k-1) - \theta(k-1)Q(k) \nonumber\\
=&\, x(k) - (I_N \otimes A) \hat x(k-1)  \nonumber\\
&- \theta(k-1)Q_R\left(\frac{x(k) - (I_N \otimes A) \hat x(k-1)}{\theta(k-1)} \right ).
\end{align}
Note that
\begin{align}\label{3}
x(k) - (I_N \otimes A ) \hat x(k-1) 
=  H  e(k-1) - L \delta (k-1)
\end{align}
where 
\begin{align}\label{HK}
H =I_N \otimes A  +  L_ \mathcal{G}\otimes B K. 
\end{align}
Then (\ref{e no DoS 1}) can be rewritten as 
\begin{align}\label{e no DoS}
e(k) 
=&\, H   e(k-1) -L \delta (k-1)    \nonumber\\
&-  \theta(k-1) Q_R \left(\frac{ H  e(k-1) -L \delta (k-1) }{\theta(k-1)} \right).
\end{align}

\emph{If the transmission fails} such that $k \in H_q$ for $k= 1, 2, \cdots$, then in view of (\ref{eq estimator all}), one has 
\begin{align}\label{e DoS}
e(k) &= x(k) - \hat x(k)=   x(k) - (I_N \otimes A) \hat x(k-1). 
\end{align}
Then apply (\ref{3}) to (\ref{e DoS}). 

In the above, we have presented the system dynamics using $e(k)$ and $\delta(k)$.
To facilitate the analysis, we let
\begin{align}\label{transformation}
\alpha(k) = \delta(k)/\theta(k)   \quad \quad
\xi(k) = e(k)/\theta(k) 
\end{align}
where $\theta(k)$ is given in (\ref{eq h}). Then we formulate the system dynamics in terms of $\alpha(k)$ and $\xi(k)$.

\emph{If the transmission succeeds} such that $k \notin H_q$, in view of the first relation in (\ref{eq h}), (\ref{delta all}) and (\ref{e no DoS}), one has
\begin{align}
\alpha(k) =& \frac{ G}{\gamma _ 1} \alpha(k-1) + \frac{L}{\gamma _ 1} \xi(k-1) \label{alpha no DoS} \\
\xi(k)=& \frac{ H \xi(k-1) -L \alpha(k-1)   }{\gamma_1}  \nonumber\\
& -  \frac{ Q_R \left( H \xi(k-1) - L \alpha(k-1)\right)}{\gamma_1}.   \label{xi no DoS}
\end{align}
It is easy to infer that if $\| H \xi(k-1) -L \alpha(k-1) \|_\infty \le (2R+1)\sigma $, then by (\ref{quantizer error}) one has $\|\xi(k)\|_\infty \le \sigma / \gamma_1$.

\emph{If the transmission fails} such that $k\in H_q$, then according to the second case in (\ref{eq h}), (\ref{delta all}) and (\ref{e DoS}), one has
\begin{align} 
\alpha(k) &= \frac{G }{\gamma _ 2} \alpha(k-1) + \frac{L}{\gamma _ 2} \xi(k-1) \label{alpha DoS}   \\ 
\xi(k) &= \frac{H}{\gamma_2} \xi(k-1) - \frac{ L }{\gamma_2} \alpha(k-1).  \label{xi DoS} 
\end{align}
Compared with (\ref{xi no DoS}), $\xi(k)$ induced by (\ref{xi DoS}) may not satisfy $\|\xi(k)\|_\infty \le \sigma / \gamma_1$. In the event that $\|\xi(k)\|_\infty > \sigma / \gamma_1$, there is a possibility that $\|H  \xi(k) - L \alpha(k) \| _\infty > (2R+1)\sigma$, which demonstrates that quantizer overflow occurs. 

We explain the intuition of the zooming-in and zooming-out mechanism in the context of quantized control of multi-agent systems under transmission losses. 
In the dynamics of $\alpha(k)$ and $\xi(k)$ in (\ref{alpha no DoS}) and (\ref{xi no DoS}) under successful transmissions, one can see that $\gamma_1$ appears in the denominators on the right-hand sides. 
Similarly, in (\ref{alpha DoS}) and (\ref{xi DoS}), $\gamma_2$ appears in the case of transmission failures. 
When DoS attacks occur, 
the systems are in the open-loop status
and thus $\alpha(k)$ and $\xi(k)$ grow in general. The parameter $\gamma_2$ can be considered as a factor to compensate the growth rate.
To keep the growth of $H  \xi(k) - L \alpha(k)$ small, we must find
a sufficiently large $\gamma_2$ since
$\alpha(k)$ and $\xi(k)$ are divided 
by $\gamma_2$ during DoS (see the right-hand sides of (\ref{alpha DoS}) and (\ref{xi DoS})).
As a result, it is possible to keep
$\|H  \xi(k) - L \alpha(k) \| _\infty \le (2R+1)\sigma$
during DoS, which implies
that quantizer overflow will not occur.

While the idea of zooming-in and zooming-out is intuitive, the computation of the parameters $\gamma_1$ and $\gamma_2$ is not straightforward in the context of quantized control of multi-agent systems. Compared with quantized control of centralized systems, e.g. in \cite{feng2020tac,867021}, 
one of the challenges in this paper arises from the constraint of distributed systems, where 
each agent knows only a fraction of the global information. Due to this, the ``decedent" state estimation/prediction scheme as in the papers \cite{feng2020tac,867021} is very difficult to implement here and more importantly the estimation error is also coupled with the state, e.g. $\xi(k)$ depends on $\alpha(k)$ in (\ref{xi DoS}). By contrast, in quantized control of centralized systems, this coupling problem between estimation error and state does not arise.

In the following, with the control scheme introduced in (\ref{controller}) to (\ref{eq h}), we will show that quantizer overflow will not occur by properly designing the scaling parameter $\theta(k)$ in (\ref{eq h}) with $\gamma_1$ and $\gamma_2$, and then discuss the trade-offs between resilience and data rate.

\subsection{Overflow-free quantizer and leaderless consensus}
In this subsection, we will present the results for quantized leaderless consensus under DoS, showing the number of quantizer levels such that it is not overflowed, and a sufficient condition for consensus. Before presenting the results, we introduce some preliminaries that will be used in the theorem.

Using the matrices $G$, $L$ and $H$ in (\ref{G}) and (\ref{HK}), respectively, we define the matrices
\begin{align}\setlength{\arraycolsep}{1.9pt} \label{bar A}
\overline A = \left[
\begin{array}{rr}
G & L \\
-L & H 
\end{array}\right], \,
\overline A(m) = \overline A ^ m = 
\left[\begin{array}{rr}
\overline A_{11}(m) & \overline A_{12}(m)  \\
\overline A_{21}(m)  & \overline A_{22}(m)  
\end{array}\right]
\end{align}
where $\overline A_{11}(m), \overline A_{12}(m), \overline A_{21}(m)$ and $\overline A_{22}(m)$ are compatible submatrices with dimensions $nN \times nN$ in $\overline A(m) $ and the integer $m$ satisfies $0 \le m \le M $ as in Lemma \ref{Lemma Q}.
Then, we define $G(m+1)$ and $\overline G (m+1)$ as
\begin{align}
G(m+1) &=  (G \overline A_{11}(m) + L \overline A_{21}(m))/\gamma_2^m \label{G(m+1)} \\
\overline G (m+1) &= (U \otimes I_n)^T G(m+1) (U \otimes I_n) \label{bar G}
\end{align}
in which the unitary matrix $U$ is given by
\begin{align}\label{Phi}
U = [\mathbf{1}/ \sqrt{N} \,\, \phi_2 \,\, \cdots \,\, \phi_N  ] \in \mathbb{R}^{N \times N}
\end{align}
where $\phi_i \in \mathbb{R}^N$ with $i=2, 3, \cdots, N $ satisfies $\phi^T _i L_\mathcal{G} = \lambda_i \phi_i ^ T $.
Let the matrix $J(m+1) \in  \mathbb{R} ^{n(N-1) \times n(N-1) }$ denote the remaining parts of $\overline G(m+1)$ in (\ref{bar G}) after deleting the top $n$ rows and the left $n$ columns from $\overline G(m+1)$. Then we define the set $ \mathcal J $ as
\begin{align}\label{J}
\mathcal J  =\{J(1),  \cdots, J(m+1), \cdots, J(M+1)\}. 
\end{align} 
Note that $J(m+1)$ is reduced to $J(1)$ in (\ref{J(1)}) when $m=0$, which is independent of $\gamma_2$. If $1 \le m \le M$, then $J(m+1)$ is dependent on $\gamma_2$. With the matrices $\overline A_{12}(m)$ and $\overline A_{22}(m)$ in (\ref{bar A}), and $G$ and $L$ in \eqref{G}, we let 
\begin{align}\label{L(m+1)}
L(m+1) =  ( G \overline A_{12}(m) + L \overline A_{22}(m)) / \gamma_2^m
\end{align}
and then compute 
\begin{align}\label{C5}
C_0 = \max_{m=0, 1, \cdots, M} \|L(m+1)\|.
\end{align}
With such $C_0$, we further compute
\begin{align}\label{C6}
C_1 = \max\left\{2C_2  \sqrt{Nn} ,  \frac{C_0 C_2 \sqrt{Nn} \sigma}{ (1-d) \gamma_1  }  \right\}  \in \mathbb{R}_{>0}
\end{align}
where $d  = d_0 / \gamma_1$. Here, the parameters $\rho (J(1)) < d_0 <1 $ and $C_2 \ge 1$ in (\ref{C6}) exist and satisfy $\| J(1)  ^ p \| \le C_2 d_0 ^p$ with $p \in \mathbb Z _{\ge 1}$ \cite{horn2012matrix}. The choices and discussions concerning $\gamma_1$ and also $\gamma_2$ will be given in Lemma \ref{Lemma3} and thereafter.

To facilitate the proof of Theorem \ref{Theorem 1}, we introduce the lemma below, whose proof is provided in the Appendix. 
\begin{lemma}\label{Lemma3}
	Take $\gamma_1$ and $\gamma_2$ such that 
	\begin{align}\label{eq Lemma3}
	d_0 < \gamma_1 < 1,  \,\, \max_{m=1, \cdots, M} \|J(m+1)\| \le  \rho(J(1))/C_2 
	\end{align}
	in which $M$ is from Lemma \ref{Lemma Q} and $C_2  \in \mathbb R_{\ge 1} $ satisfies $\|J(1)^p\| \le C_2 d_0 ^p$ with $\rho(J(1)) < d_0 <1 $. Let $\theta_0 \ge C_{x_0} \gamma_1 / \sigma$. If $\|\xi(s_p)\|_\infty \le  \sigma/ \gamma_1$ for $p=0, 1, \cdots, r$, then $\| [\alpha^T(s_r) \,\,\,  \xi^T(s_r)  ]^T\|$ is upper-bounded as
	\begin{align}\label{vector Lemma 3}
	\| [\alpha^T(s_r) \,\,\,  \xi^T(s_r)  ]^T\| \le \sigma \sqrt{C_1 ^2 +  Nn}  / \gamma_1
	\end{align} 
where $C_1$ is given in (\ref{C6}).  \qedp
\end{lemma}

After finding $C_2$ and $d_0$ for $\| J(1)  ^ p \| \le C_2 d_0 ^p$, one must first choose a $\gamma_1$ such that $ d_0 < \gamma_1 < 1 $. Recall that $\gamma_2$ appears in the denominators of $J(2), \cdots, J(M+1)$ by (\ref{G(m+1)}). Then, one selects a sufficiently large $\gamma_2$ such that the second inequality in (\ref{eq Lemma3}) holds. 
Note that as long as $C_2$ and $d_0$ are determined, the choices of $\gamma_1$ and $\gamma_2$ can be made independently.

Now we are ready to present the results for quantized leaderless consensus under DoS attacks.

\begin{theorem}\label{Theorem 1}
	Consider the multi-agent system (\ref{system}) with control inputs (\ref{controller}) to (\ref{eq h}), where the agents exchange information via the undirected graph $\mathcal{G}$. The communication attempts are periodic with sampling interval $\Delta$. Suppose that the DoS attacks characterized in Assumptions 1 and 2 satisfy $1/T + \Delta/\tau_D < 1$. 
	The parameters $\gamma_1$, $\gamma_2$ and $\theta_0$ are chosen in accordance to Lemma \ref{Lemma3}. If $R$ satisfies
	\begin{align}\label{Theorem quantizer}
	2R+ 1 
	\ge  \|[-L   \,\, H  ] \|_\infty \zeta \frac{\sqrt{C_1 ^ 2 +  Nn}}{\gamma_1}
	\end{align}
	with $C_1$ in (\ref{C6}), $\zeta =\max_{m=0, 1, \cdots, M}\|(\overline A / \gamma_2 ) ^ m  \| $, $\overline A $ in (\ref{bar A}) and $M$ in Lemma \ref{Lemma Q}, then the following hold:
	\begin{itemize}
		\item[\textbf{i)}] The quantizer (\ref{quantizer}) is not overflowed. 
		\item[\textbf{ii)}] 	If in addition the DoS attacks satisfy
		\begin{align}\label{Theorem DoS}
		\frac{1}{T} + \frac{\Delta}{\tau_D} < \frac{-\ln \gamma_1}{\ln \gamma_2 - \ln \gamma_1},
		\end{align}
		then consensus of $x_i(k\Delta)$ is achieved as in (\ref{control objective}). 
	\end{itemize}

\end{theorem}
\color{black}


\emph{Proof.} Recall that $s_r$ represents a successful transmission instant, In the proof, we let $s_r$ represent $k \notin H_q$. 

\textbf{i)} The unsaturation of the quantizer is proved by induction. More specifically we show that if the quantizer is not overflowed such that $\|\xi (s_r)\|_\infty \le \sigma / \gamma_1 $ for $r=-1, 0, \cdots$, then the quantizer will not saturate at the transmission attempts within the interval $]s_r, s_{r+1}]$ and hence $\|\xi (s_{r+1})\|_\infty \le \sigma / \gamma_1 $.

\emph{a)} If $s_{r+1}=s_{r} + \Delta$, in view of (\ref{xi no DoS}), it is easy to verify that the quantizer $Q(s_{r+1}) = Q_R( H\xi(s_{r})-L \alpha(s_{r} )  )   $ is not overflowed in the sense that 
\begin{align}
\left\| [-L \quad H\,]   
\left[
\begin{array}{ll}
\alpha^T(s_r) &
\xi^T (s_r)
\end{array}
\right]^T
\right\|_\infty 
\le (2R + 1) \sigma
\end{align}
where the norm of $[\alpha^T(s_r) \,\, \xi^T(s_r)]^T$ is given in Lemma \ref{Lemma3}. This implies $\| \xi (s_{r+1}) \| _\infty \le \sigma / \gamma_1$.

\emph{b)} If $s_{r+1} > s_{r} + \Delta$, it means that the transmissions before $s_{r+1}$ at the instants $s_{r}+ \Delta$, $s_{r}+ 2\Delta$, $\cdots, s_{r} + m\Delta$ fail, where $ m \le M$. We verify that the quantizer is also free of overflow at the instants $s_{r}+ \Delta$, $s_{r}+ 2\Delta$, $\cdots, s_{r} + m\Delta$ and $s_{r+1}$ since
\begin{align}
&\left\| [-L \,\,\, H\,]   
\left[
\begin{array}{ll}
\alpha^T(s_{r}+m\Delta) &
\xi^T (s_{r}+m\Delta)
\end{array}
\right]^T
\right\|_\infty \nonumber\\
&\le 
\| [-L \,\,\, H\,] \|_\infty
\left\|\overline A(m)/\gamma_2^m \right\|
\left\|
\left[
\begin{array}{ll}
\alpha^T(s_{r}) &
\xi ^T (s_{r})
\end{array}
\right]^T
\right\| \nonumber\\
&\le  \,(2R+1) \sigma, \,\,\,   0 \le m \le M.
\end{align}
This implies $\|\xi (s_{r+1}) \| _\infty  \le \sigma /\gamma_1$. In view of \emph{a)} and \emph{b)}, by induction, we conclude that the quantizer satisfying (\ref{Theorem quantizer}) is not overflowed for all transmissions. 

\textbf{ii)} Now we show leaderless consensus in the states. If the quantizer is not saturated, then one has
\begin{align}\label{alpha sr m}
\|\alpha (s_r + m \Delta)\|_ \infty 
&\le \|[\alpha^T (s_r + m \Delta)  \,\,\,  \xi^T (s_r + m \Delta)]^T\| \nonumber\\
&\le  
\left\|\overline A(m)/ \gamma_2^m \right\|
\left\|
[
\alpha^T (s_{r}) \quad 
\xi^T (s_{r})
]^T
\right\|  \nonumber\\
&\le  \sigma \left\| \overline A(m)/\gamma_2^m  \right\| \sqrt{C_1 ^ 2 +  Nn} /\gamma_1
\end{align}
for $1 \le m\le M$, where the third inequality is obtained from (\ref{vector Lemma 3}). 
Incorporating the scenario of $m=0$, we have  
$
\|\alpha (k)\|_ \infty    \le \sigma \zeta \sqrt{C_1 ^ 2 + Nn} /\gamma_1 
$
where $\zeta =\max_{m=0, 1, \cdots, M}\|(\overline A / \gamma_2 ) ^ m  \|$.
Recall the definition of $T_S(\Delta, k\Delta)$ in Lemma \ref{Lemma T} and let $T_U(\Delta, k\Delta)$ denote the number of unsuccessful transmissions in the interval $[\Delta, k \Delta]$. In view of $\delta(k) = \theta(k) \alpha(k)  = \gamma_1 ^ {T_S(\Delta,k\Delta)} \gamma_2 ^{T_U(\Delta,k\Delta)}  \theta_0 \alpha(k) $, one has
\begin{align}
\|\delta(k)\|_\infty \le  C_3  \gamma^k  \theta_0 \|\alpha(k)\|_\infty \le C_3  \gamma^k  \theta_0 \zeta \sqrt{C_1 ^ 2 +  Nn} \sigma / \gamma_1  \nonumber
\end{align}
where $C_3 = \left( \gamma_2 /\gamma_1   \right)^{(\kappa+\eta\Delta)/\Delta} $ and 
\begin{align}
\gamma = \gamma_1^{1-\frac{1}{T}-\frac{\Delta}{\tau_D}  } \gamma_2^{ \frac{1}{T}+ \frac{\Delta}{\tau_D}   } < 1
\end{align}
by (\ref{Theorem DoS}). Thus, we have $\|\delta(k)\|_\infty \to 0$ when $k\to \infty$, which implies that leaderless consensus is achieved. 
\qedp

\begin{remark}
	As mentioned earlier,
	the theorem characterizes the tradeoff between resilience of the
	agent system to DoS attacks and the necessary data rate in communication.
	This can be seen from the roles that the parameters $\gamma_1$ and $\gamma_2$
	play in our design.
	They determine the update of the scaling parameter $\theta(k)$ depending
	on the presence of DoS attacks.
	For improving the robustness, it helps to use small $\gamma_1$ and $\gamma_2$ in (\ref{eq Lemma3}),
	which will enlarge the class of tolerable DoS attacks as seen in (\ref{Theorem DoS}).
	On the other hand, a small $\gamma_1$ ($\gamma_1 \to d_0$) will result in large data rate.
	We can confirm this
	in the lower bound for $2R+1$ in (\ref{Theorem quantizer}) and also the definition of $C_1$ in (\ref{C6}).
	Intuitively, this tradeoff has a clear implication: Higher resilience needs more data rate.
	\qedp
\end{remark}
\begin{remark}\label{remark 4}
	
	Another aspect of $\gamma_1$ and $\gamma_2$ is that
	keeping them small helps the convergence rate for arriving at
	consensus. This can be checked as follows: Small $\gamma_1$ and $\gamma_2$ help the convergence rate of $\theta(k)$. Then from (\ref{transformation}), this can result in a fast convergence rate of $\delta(k)$ and hence the state consensus. Though the analysis methods in our paper and \cite{feng2019secure} are different, they have some common points. For example, it is good to have fast consensus rate by controller design during DoS-free periods. In our paper, this can be realized by tuning $K$ and enlarging data rates. In \cite{feng2019secure}, this can be realized by tuning the solution to algebraic Riccati equation. 
	\qedp
\end{remark}

Note that from the
iteration (\ref{alpha no DoS}) and the discussion after (\ref{delta all}), when DoS is absent, the
iteration of $\alpha(k)$ depends on
$J(1)/\gamma_1$,
which is similar to the result achieved in \cite{you2011network}.
In this paper, we extend this to the case when $0\le m \le M$ consecutive packet
dropouts can occur, where the condition is written in terms of
$J(m+1)/\gamma_1$ with $J(m+1)$ in (\ref{eq Lemma3}). The data rate result given by (\ref{Theorem quantizer}) can be conservative than the corresponding one in \cite{you2011network}. This is due to the worst-case type of analysis when considering uncertain DoS attacks, e.g., the use of $\max$ in $C_0$ in (\ref{C5}) and $\zeta$ in (\ref{Theorem quantizer}). At last, the parameter $C_2$ in (\ref{eq Lemma3}) can also make $\gamma_2$ conservative. The purpose of letting $C_2$ be in the denominator in (\ref{eq Lemma3}) is for compensating the ``jumps" in the switched system from one mode to the others (DoS modes and non-DoS mode).

It is clear from our results that a control designer may not need the exact knowledge of the real-time DoS parameters. He/She only
needs to assume that DoS attacks satisfy the condition $1/T + \Delta/ \tau_D<1$, under which Lemmas \ref{Lemma Q} and \ref{Lemma T} hold.
At or above the threshold 1 (i.e., $ 1/T + \Delta/ \tau_D \ge 1$), $\tau_D$ and $T$ can give rise to DoS signals that destroy consensus, no matter what the controller is, e.g., $T=1$ (DoS attacks are present for 100\% of total time) and/or $\tau_D=\Delta$ (DoS attacks can coincide with all transmission instants).
Furthermore, a designer may estimate the DoS parameters ($\eta$, $\tau_D$, $\kappa$ and $T$) from past experience and may also add safety margins to the parameters to ensure more robustness in the design.


\section{Leader-follower consensus under DoS }

In this section, we will discuss leader-follower consensus under DoS attacks.
The dynamics of the followers is taken as (\ref{system}). Let $0$ be the index for the leader. 
The dynamics of the leader is given as an autonomous system such that 
\begin{align}\label{leader}
x_0(k\Delta) = A x_0((k-1)\Delta), \,\, k \in \mathbb{Z}_{\ge 1}
\end{align}
where $x_0(k) \in \mathbb{R}^n$ is the state of the leader, and $A$ and $\Delta$ are the same as in (\ref{system}). Similarly to the scenario of leaderless consensus, we assume that an upper bound on the initial state of the leader is known as $\|x_0(0)\| _ \infty \le \widetilde C_{x_0}$. For the ease of analysis, we assume that $\widetilde C_{x_0} \le C_{x_0}$. 
We say that the leader-follower consensus is achieved if
\begin{align}\label{LF}
\lim_{k  \to \infty } \| x_i(k\Delta) -   x_0 (k\Delta)  \| _ \infty = 0, \,\,\, i=1,2,\cdots, N.
\end{align}

\textbf{Communication topology.}
In this section, the communication topology among the followers is represented by an undirected and connected graph $\mathcal G$ as in Section II--A, whose Laplacian matrix is denoted by $L_\mathcal{G}$.
We also assume that only a fraction of the followers can receive the information from the leader. Let $a_{i0}$ represent the leader-follower interaction, i.e. if agent $i$ can directly receive the information from the leader, then $a_{i0}>0$, and otherwise $a_{i0}=0$. Moreover, we let the diagonal matrix be $D= \text{diag}(a_{10}, a_{20}, \cdots, a_{N0}) \in \mathbb{R}^{N \times N}$. For simplicity, we let $k$ represent $k\Delta$ in the following analysis.

\subsection{Framework of leader-follower control }
For achieving the leader-follower consensus as in (\ref{LF}), we let the control input to the follower agent $i \in \mathcal V$ in (\ref{system}) as 
\begin{align}\label{lf u}
u_i (k) &= K \sum_{j=1}^{N}a_{ij} (\hat x_j(k) - \hat x_i(k)) + K a_{i0} (\hat x_0 (k) - \hat x_i (k)) 
\end{align}
where $\hat x _ j (k) $ denotes the estimate of $x_j(k)$ obtained by (\ref{eq estimator}) and (\ref{Q_i}) for $j \in \{i\} \cup \mathcal{N}_i $. Besides, $\hat x_0 (k)$ denotes the estimation of $x_0(k)$ and is also estimated as in (\ref{eq estimator}) and (\ref{Q_i}). 
The zooming-in and zooming-out quantization mechanism is still valid for leader-follower consensus control. 
The scaling parameter $\theta(k)$ is in the form as in (\ref{eq h}). 
The zooming-in and zooming-out parameters $\gamma_1$ and $\gamma_2$ for leader-follower consensus will be given later in this section.    
Here we assume that there exists a feedback gain $K \in \mathbb{R} ^{w \times n}$ for leader-follower consensus such that the spectral radius of $A- \widetilde \lambda_i BK$ ($i=1, 2, \cdots, N$) are smaller than 1, 
where $\widetilde \lambda _i $ denote the eigenvalues of $L_{\mathcal G} + D$.
We let $\widetilde \delta_i (k) = x_i(k) - x_0(k)$ and $e_i(k) = x_i(k) - \hat x_i (k)$. Moreover, let $e_0(k) = x_0(k) - \hat x_0 (k)$.
Let the vectors be $\widetilde \delta (k)  = [\widetilde \delta_1 ^ T (k) \,\,\, \widetilde  \delta_2 ^ T (k)\,\,\, \cdots  \widetilde \delta_N ^ T (k)]^T $ and $e (k) = [e_1 ^ T (k) \,\,\,  e_2 ^ T (k)\,\,\, \cdots  e_N ^ T (k)]^T $. 
Then we obtain the compact form
\begin{align} \label{lf delta}
\widetilde \delta (k) = \Pi \widetilde \delta(k-1) + \Sigma  e(k-1) - \Phi (1_N \otimes e_0(k-1))
\end{align}
where the matrices are given by 
$
\Pi = I_N \otimes A -(L_{\mathcal{G}}+D) \otimes BK,\,\, \Sigma = (L_{\mathcal G}+D) \otimes BK$ and $ 
\Phi  = D \otimes BK.
$
Note that the eigenvalues of $\Pi$ equal to those of 
$A- \widetilde \lambda_i BK$ with spectral radius $\rho(A- \widetilde \lambda_i BK)<1$ ($i=1, 2, \cdots, N$). 
If the dynamics of $\widetilde \delta(k)$ is stable as $\|\widetilde \delta(k)\|_\infty \to 0$ ($k \to \infty$), then the leader-follower consensus is achieved as in (\ref{LF}). 

\subsection{System dynamics of leader-follower consensus under DoS}
In light of (\ref{lf delta}), one sees that the convergence of $\widetilde \delta(k)$ depends on $e(k)$ and $e_0(k)$. We first analyze $e_0(k)$, whose dynamics follows
\begin{align}
e_0(k) = 
\left\{
\begin{array}{ll}
A e_0 (k-1) - \theta (k-1)  Q_R \left(\frac{A e_0 (k-1)}{\theta(k-1)}\right)  & k \notin H_q\\
A e_0 (k-1) &  k \in H_q.
\end{array}\right.
\end{align}
It is clear that the dynamics of $e_0(k)$ depends on only $e_0(k-1)$, which is different from that in leaderless consensus where the dynamics of $e_i(k)$ depends on $e_i(k-1)$, $e_j(k-1)$, $\delta_i(k-1)$ and $\delta_j(k-1)$ ($j\in \mathcal N_i$). This is because that the leader agent does not receive information from its neighbors and hence its state is decoupled from those of the followers. On the other hand, the phenomenon that the estimation errors of followers' states are still coupled as occurred in the leaderless consensus problem. As we will see later, the estimation errors of followers' states are also coupled with $e_0(k)$.

Now we discuss the evolution of $e(k)$.
In the scenario of leader-follower consensus, the equations (\ref{e no DoS 1}) and (\ref{e DoS}) still hold. However, the item $x(k) -( I_N \otimes A )\hat x(k-1)$ is different from the one in (\ref{3}), and now it is in the form of
\begin{align}\label{4}
&\, x(k) -( I_N \otimes A )\hat x(k-1) \nonumber\\
& = \, \Omega e(k-1) - \Sigma \widetilde \delta(k-1) - \Phi (1_N \otimes e_0(k-1)  )
\end{align}
where
$
\Omega = I_N \otimes A + (L _ \mathcal{G} + D) \otimes BK.
$
Substituting (\ref{4}) into (\ref{e no DoS 1}) and (\ref{e DoS}), respectively, one can obtain the dynamics of $e(k)$ in the absence and presence of DoS attacks in the scenario of leader-follower consensus. Due to space limitation, we omit presenting them. Define three vectors $\beta(k), \epsilon(k)$ and $\epsilon_0(k) \in \mathbb R^{nN}$
\begin{align}
\beta (k) = \frac{\widetilde \delta(k)}{\theta(k)}, \,\, \epsilon(k) = \frac{e(k)}{\theta(k)}, \,\, \epsilon_0(k)= \frac{1_N \otimes e_0(k)}{\theta(k)}.
\end{align}
Then we obtain the dynamics of these variables for the two cases, i.e., successful and failed transmissions.

\emph{If the transmission succeeds} such that $k \notin H_q$, we have 
\begin{align}
&\beta(k) =  \, \frac{\Pi}{\gamma_1} \beta(k-1) + \frac{\Sigma}{\gamma_1} \epsilon(k-1) - \frac{\Phi}{\gamma_1} \epsilon_0(k-1) \label{beta no DoS} \\
&\epsilon(k) =  \, \frac{\Omega }{\gamma_1} \epsilon(k-1) - \frac{\Sigma}{\gamma_1} \beta(k-1) - \frac{\Phi}{\gamma_1}  \epsilon_0(k-1) \label{epsilon no DoS}		\nonumber\\
&\quad\quad - \frac{1}{\gamma_1} Q_R \left(  \Omega\epsilon(k-1) - \Sigma \beta(k-1)  -  \Phi \epsilon_0(k-1) \right)  \\
&\epsilon_0(k) =  \, \frac{I_N \otimes A}{\gamma_1} \epsilon_0(k-1) - \frac{1}{\gamma_1}  Q _R ((I_N \otimes A) \epsilon_0(k-1)). \label{epsilon 0 no DoS}
\end{align}
\emph{If the transmission fails} such that $k \in H_q$, we have
\begin{align}
\beta(k) &= \frac{\Pi}{\gamma_2} \beta(k-1) + \frac{\Sigma}{\gamma_2} \epsilon(k-1) - \frac{\Phi}{\gamma_2}  \epsilon_0(k-1) \label{beta}		\\
\epsilon(k) &= \frac{\Omega }{\gamma_2} \epsilon(k-1) - \frac{\Sigma}{\gamma_2} \beta(k-1) - \frac{\Phi}{\gamma_2} \epsilon_0(k-1)  \label{epsilon}   \\ 
\epsilon_0(k) &= \frac{I_N \otimes A}{\gamma_2} \epsilon_0(k-1).  \label{epsilon 0}
\end{align}


Comparing the expressions of $Q_R(\cdot)$ in (\ref{xi no DoS}) and (\ref{epsilon no DoS}), one sees that the dynamics of $\epsilon(k)$ (transformed estimation error of follower state) also depends on $\epsilon_0(k)$ (transformed estimation error of leader state).
By contrast, in the leaderless consensus problem, this does not occur. Therefore, the leader state also needs be properly quantized. This is one of the major differences of leader-follower consensus from the leaderless one. By (\ref{epsilon no DoS}) and (\ref{epsilon 0 no DoS}), it is easy to infer that if $\| \Omega\epsilon(k-1) - \Sigma \beta(k-1)  -  \Phi \epsilon_0(k-1)  \|_\infty \le (2R+1)\sigma $ and $\|(I_N \otimes A) \epsilon_0(k-1)\|_\infty \le (2R + 1) \sigma$, then by (\ref{quantizer error}) one has $\|\epsilon(k)\|_\infty \le \sigma / \gamma_1$ and $\|\epsilon_0(k)\|_\infty \le \sigma / \gamma_1$, respectively. This means that if the transmissions succeed at $k$, $\epsilon(k)$ and $\epsilon_0(k)$ can be reset.

By (\ref{epsilon}), it is possible that $\| \epsilon(k)\|_\infty \le \sigma / \gamma_1$ does not hold during DoS, since $\epsilon(k)$ cannot be reset as in (\ref{epsilon no DoS}). Similar to the case in the leaderless consensus problem, 
here in the event that $\| \epsilon(k)\|_\infty > \sigma / \gamma_1$, there is also a possibility that $\| \Omega\epsilon(k) - \Sigma \beta(k)  -  \Phi \epsilon_0(k)  \| _\infty > (2R+1)\sigma$, which demonstrates that quantizer overflow for the follower state occurs. Moreover, in view of (\ref{epsilon 0 no DoS}) and (\ref{epsilon 0}), the overflow problem can also happen to the quantization of leader state during DoS. In the following, with the control scheme introduced in (\ref{lf u}), we will show that quantizer overflow for both leader and follower states will not occur if one properly designs the scaling parameter $\theta(k)$ in (\ref{eq h}). Then we will discuss the trade-offs between resilience and data rate.

\subsection{Result for leader-follower consensus}
To facilitate the subsequent analysis of leader-follower consensus, we introduce some preliminaries.

In view of the matrices $\Pi$, $\Sigma$, $\Phi$ and $\Omega$ in (\ref{lf delta}) and (\ref{4}), respectively, we define the matrices
\begin{align}\setlength{\arraycolsep}{1.9pt} 
\widetilde A &= \left[\begin{array}{rrr}
\Pi &\quad \Sigma & -\Phi \\
-\Sigma &\quad  \Omega & - \Phi \\
\mathbf 0 & \mathbf 0  & I_N \otimes  A
\end{array}\right] \label{tilde A} \,\,\, \text{and} \\
\widetilde A (m) =\widetilde A ^ m 
& =
\left[\begin{array}{rrr}
\widetilde A _{11}(m) &  \widetilde A _{12}(m) & \widetilde A _{13}(m)\\
\widetilde A _{21}(m) &\widetilde A _{22}(m)  & \widetilde A _{23}(m) \\
\mathbf 0 &  \mathbf 0 & I_N \otimes  A^m
\end{array}\right] \label{tilde Am}
\end{align}
where $\widetilde A _{11}(m)$, $\widetilde A _{12}(m)$, $\widetilde A _{13}(m)$, $\widetilde A _{21}(m)$, $\widetilde A _{22}(m)$ and $\widetilde A _{23}(m)$ are compatible submatrices of $\widetilde A(m)$ and the integer $m$ satisfies $0 \le m \le M $ as in Lemma \ref{Lemma Q}.
Then, we define $
P(m+1) =  (\Pi \widetilde A_{11} (m) + \Sigma \widetilde A_{21}(m))/\gamma_2 ^m$, 
$S(m+1)  =  ( \Pi \widetilde A_{12} (m) + \Sigma \widetilde A_{22}(m) )/ \gamma_2 ^m$
and 
$Z(m+1) = ( \Pi \widetilde A_{13}(m) + \Sigma \widetilde A_{23}(m)  - \Phi (I_N \otimes A^m) ) / \gamma_2 ^m$. Let $\widetilde C_0 = \max_{m=0, 1, \cdots, M} \|S(m+1)\|$ and $\widetilde C_1 = \max_{m=0, 1, \cdots, M} \|Z(m+1)\|.$
There exists a unitary matrix $\widetilde \Psi$ such that $\widetilde \Psi ^ {-1} (L _{\mathcal G} + D) \widetilde \Psi$ is an upper-triangular matrix whose diagonals are $\widetilde \lambda_i$ ($i=1, 2, \cdots, N$), which are the eigenvalues of $L _{\mathcal G} + D$ \cite{horn2012matrix}. With the $\widetilde \Psi$, we define the matrices
\begin{align}\label{tilde P}
\widetilde P(m+1)  = (\widetilde \Psi \otimes I_n)^T P(m+1)(\widetilde \Psi \otimes I_n).
\end{align}
Then we define the set of matrices $\mathcal P$ as 
\begin{align}\label{set P}
\mathcal P = \{ \widetilde P(1),  \cdots, \widetilde P(m+1), \cdots, \widetilde P(M+1)\}
\end{align}
where in particular we have
\begin{align}\label{tilde P1}\setlength{\arraycolsep}{1.5pt}
\widetilde P(1) = 
\left[
\begin{array}{cccc}
A- \widetilde \lambda_1 BK & \star & \star   & \star \\
\mathbf 0					   & A- \widetilde \lambda_2 BK & \star & \star \\
\mathbf 0		&\mathbf 0		 & \ddots &  \vdots \\
\mathbf 0		&\mathbf 0		&\mathbf 0		& A- \widetilde \lambda_N BK
\end{array}
\right].
\end{align}
with $\star$ presenting compatible matrices. 
Finally, we let 
\begin{align}\label{C9}
\widetilde C_3 = \max \left\{	2 \widetilde C_4 \sqrt{Nn}, \frac{\widetilde C_2  \widetilde C_4 \sqrt{Nn}}{(1-\widetilde d)\gamma_1}	\right\}
\end{align}
where $\widetilde C_2 = \widetilde C_0 + \widetilde C_1$. The parameters $\tilde d = \tilde d_0 /\gamma_1 $ and $\widetilde C_4$ are obtained from the following computation $\rho (\widetilde P(1)) < \tilde d_0<1$ and $\widetilde C_4 \ge 1$ satisfying $\|\widetilde P (1) ^ k\| \le \widetilde C_4  \tilde d_0 ^ k $ with $k \in \mathbb Z _{\ge 1}$.

To facilitate the proof, we first present the following lemma whose proof is provided in the Appendix. 

\begin{lemma}\label{Lemma4}
	Take $\gamma_1$ and $\gamma_2$ such that 
	\begin{align}\label{eq Lemma4}
	\tilde d_0 < \gamma_1 < 1,  \,\, \max_{m=1, \cdots, M} \|\widetilde P (m+1)\| \le  \rho(\widetilde P(1))/\widetilde C_4 
	\end{align}
	in which $M$ in Lemma \ref{Lemma Q}, and $\tilde C_4  \ge  1$ satisfying $\|\widetilde P(1)^ p\| \le \widetilde C_4 \tilde d_0 ^p$ with $ \rho(\widetilde P (1)) < \widetilde d_0  < 1$. Let $\theta_0 \ge C_{x_0} \gamma_1 / \sigma$. If $\| \epsilon (s_p) \|_\infty  \le \sigma  / \gamma_1$ and $\|\epsilon_0(s_p)\| _ \infty \le \sigma / \gamma_1$ for $p=0, \cdots, r$, then $\|[\beta ^T (s_r)\,\,\, \epsilon^T(s_r)\,\,\,   \epsilon_0 ^ T (s_r)  ] ^ T\|$ is upper-bounded as
	\begin{align}\label{vector Lemma4}
	\|[\beta ^T (s_r)\,\,\, \epsilon^T(s_r)\,\,\,   \epsilon_0 ^ T (s_r)  ] ^ T\| \le \sigma \sqrt{\widetilde C_3 ^ 2 + 2Nn}/\gamma_1
	\end{align}
	with $\tilde C_3$ in (\ref{C9}).   \qedp
\end{lemma}

Now we are ready to present the leader-follower results.

\begin{theorem}\label{Theorem 2}
	Consider the multi-agent system (\ref{system}) as the follower agent with control action (\ref{lf u}), (\ref{eq estimator}) to (\ref{eq h}). The leader agent is given in (\ref{leader}). The communication attempts are periodic with sampling interval $\Delta$. Suppose that the DoS attacks in Assumptions 1 and 2 satisfy $1/T + \Delta/\tau_D < 1$.
	Let $\theta_0$, $\gamma_1$ and $\gamma_2$ be chosen as in Lemma \ref{Lemma4}. If $R$ satisfies
	\begin{align}\label{lf Theorem quantizer}
	2R+ 1 
	\ge \widetilde \zeta  \| [-\Sigma \quad \Omega \quad  -\Phi ] \|_\infty  \sqrt{\widetilde C_3 ^ 2 + 2Nn} /\gamma_1
	\end{align}
	with bounded reals $\widetilde \zeta = \max \{  \widetilde \zeta_1,\widetilde \zeta_2\} $ and $\widetilde C_3 \in \mathbb{R}_{>0}$ in (\ref{C9}), then the following hold: 
	\begin{itemize}
		\item[\textbf{i)}] 	The quantizer (\ref{quantizer}) is not overflowed.
		\item[\textbf{ii)}] If in addition the DoS attacks satisfy (\ref{Theorem DoS}), then the leader-follower consensus as in (\ref{LF}) is achieved. 
	\end{itemize}

	
\end{theorem}
\color{black}

\emph{Proof.} 
\textbf{i)} The unsaturation of the quantizer is proved by induction. Specifically, if the quantizer is not overflowed such that $\|\epsilon (s_r)\|_\infty \le \sigma / \gamma_1 $ and $\|\epsilon_0(s_r)\|_ \infty \le \sigma/ \gamma_1$ for $r=-1, 0, \cdots$, then the quantizer will not saturate at the transmission attempts within $]s_r, s_{r+1}]$, which implies $\|\epsilon (s_{r+1})\|_\infty \le \sigma / \gamma_1 $ and $\|\epsilon_0(s_{r+1})\|_\infty \le \sigma / \gamma_1$.

\emph{a)} If $s_{r+1}=s_{r} + \Delta$, in view of (\ref{epsilon no DoS}), it is easy to verify that the quantizer $ Q_R \left(  \Omega\epsilon(s_r) - \Sigma \beta(s_r)  -  \Phi \epsilon_0(s_r) \right) $ of the follower agents is not overflowed in the sense that 
\begin{align}\label{DoS follower quantizer 2}
& \|[-\Sigma \quad \Omega \quad  -\Phi ]\,\,  [\beta ^T (s_r)\,\,\, \epsilon^T(s_r)\,\,\, \epsilon_0 ^ T (s_r)  ] ^T   \| _ \infty   \le   (2R+1)\sigma \nonumber
\end{align}
by applying the bound in (\ref{vector Lemma4}) of Lemma \ref{Lemma4}. This implies $\| \epsilon (s_{r+1}) \| _\infty \le \sigma / \gamma_1$ in view of (\ref{epsilon no DoS}). It is clear that $\|A\|_\infty \le \| [-\Sigma \quad \Omega \quad  \Phi ] \|_\infty $ and $\|\epsilon_0(s_r)\|_\infty \le \sigma /\gamma_1 $. Thus, in view of (\ref{epsilon 0 no DoS}), $Q _R ((I_N \otimes A) \epsilon_0(s_r))$ for the leader state is not saturated because
\begin{align}
\|(I_N \otimes A) \epsilon_0(s_r) \|_\infty & \le \|A\|_\infty \sigma /\gamma_1 \nonumber\\
& \le \| [-\Sigma \,\, \Omega \,  -\Phi ] \|_\infty  \sigma /\gamma_1   \le (2R + 1) \sigma. \nonumber
\end{align}

\emph{b)} If $s_{r+1} > s_{r} + \Delta$, it means that the transmissions at $s_{r}+ \Delta$, $s_{r}+ 2\Delta$, $\cdots, s_{r} + m\Delta$ fail, where $ m \le M$. We verify that the quantizers for the follower states are also free of overflow at those instants as well as $s_{r+1}$ since
\begin{align}\setlength{\arraycolsep}{1pt} 
& \left\| [-\Sigma \quad \Omega \quad  -\Phi ] 
\left[
\begin{array}{ccc}
\beta(s_r + m \Delta) \\
\epsilon(s_r + m \Delta) \\
\epsilon_0(s_r + m \Delta)
\end{array}\right] 
\right\|_\infty \nonumber\\
& \le  
\left\| [-\Sigma\,\, \Omega \,  -\Phi ]  \left(  \widetilde A/\gamma_2 \right) ^ m
[\beta ^T (s_r)\,\,\, \epsilon^T(s_r)\,\,\, \epsilon_0 ^ T (s_r)  ] ^T
\right\|_\infty   \nonumber\\
&\le  \,
\widetilde \zeta_1  \| [-\Sigma \quad \Omega \quad  -\Phi ] \|_\infty  \sigma \sqrt{\widetilde C_3 ^ 2 + 2Nn} /\gamma_1 \le  (2R+1) \sigma. \nonumber
\end{align}
where $\widetilde \zeta_1 = \max_{m=0, ..., M} \|(\widetilde A/\gamma_2)^m\|$. 
Similarly, we can also verify the unsaturation of the quantizer for the leader state in the sense that 
\begin{align}
& \, \|(I_N \otimes A) \epsilon_0(s_r + m\Delta) \|_\infty  \nonumber\\
& \le  \, \|(I_N \otimes A) \left(  I_N \otimes A / \gamma_2 \right)^m \epsilon_0(s_r) \| _\infty  \nonumber\\
& \le  \, \widetilde \zeta_2 \|A\|_ \infty \sigma /\gamma_1 \le (2R+1)\sigma.
\end{align}
where $\widetilde \zeta_2 = \max_{m=0, ..., M} \|(A/\gamma_2)^m\|$. 
In view of \emph{a)} and \emph{b)} above, by induction, we conclude that the quantizer satisfying (\ref{lf Theorem quantizer}) is not overflowed for all transmissions in the scenario of leader-follower consensus. 

\textbf{ii)} Following the calculation similar to that after (\ref{alpha sr m}) in the proof of Theorem \ref{Theorem 1}, one can obtain that $\|\beta(k)\|_ \infty$ is upper-bounded. When (\ref{Theorem DoS}) is satisfied, one has $\theta(k)\to 0$
and hence $\|\widetilde \delta(k)\|_\infty \to 0$ with $k\to \infty$, which implies that the leader-follower consensus in (\ref{LF}) is achieved. 
\qedp

Similar to the leaderless consensus scenario, it is good to
have small $\gamma_1$ that results in large data rate,
and small $\gamma_2$ for improving the robustness.
Here, for the ease of analysis,
we have taken the quantizers for the leader
and followers to be identical. 
If one deploys non-identical quantizers,
then there might be another trade-off
in terms of data rates.
By increasing the data rate for the
leader quantization,
more accurate estimation of $x_0(k)$
is possible. In turn, we may be able to
reduce the data rate among the followers.
By doing so, if the number
of the follower agents is not that small,
we expect that the
overall communication load can be reduced while in contrast
the resilience of the systems is not affected.
\\
For leader-follower consensus, Remark \ref{remark 4} still holds, i.e., it is good to keep $\gamma_1$ and $\gamma_2$ small, and have a fast consensus speed for DoS-free periods. For more details, we refer the readers to Remark \ref{remark 4} in our paper and Section IV in \cite{feng2019secure}.

\section{Numerical example}
In this section, we conduct simulations to verify our results. We consider eight agents in the leaderless consensus and also eight follower agents in the leader-follower consensus (i.e., $N=8$ in both cases). Each agent has four states with $A \in \mathbb R ^{4 \times 4}$ given below, whose spectral radius is $\rho(A) =1.1025$. The sampling period is given by $\Delta = 0.1$s.
\begin{align}
A=& 
\left[\begin{array}{rrrr}
1.1052  &  0.1105& -0.1& 0 \\
0   & 1.1052 & 0&0\\
0.1 & 0 & 0.25& 0.1\\
0.1 & 0.3& 0 & 0.2
\end{array}
\right],\, \nonumber\\ 
B= &
\left[\begin{array}{rrrr}
0.1052  &  0.0053\\
0   & 		0.1052\\
0& 0 \\
0& 0 
\end{array}
\right], \nonumber
\begin{array}{l}
K_1=
\left[\begin{array}{rrrr}
3  &  0& 0& 0 \\
0   & 3 & 0&0
\end{array}
\right], \\
K_2=
\left[\begin{array}{rrrr}
2.9  &  0& 0& 0 \\
0   & 2.9 & 0&0
\end{array}
\right]. 
\end{array}
\end{align}
For leaderless consensus, the eight agents exchange data through an undirected and connected communication graph $\mathcal{G}$. For leader-follower consensus, the communication topology among the followers is the same as the one in the leaderless consensus, that is $\mathcal{G}$. The leader agent has interactions with some of the follower agents, which is specified by the matrix $D$.
The matrices $L_\mathcal{G} \in \mathbb R ^{8 \times 8} $ and $D \in \mathbb R ^{8 \times 8} $ are given by
\begin{align}\setlength{\arraycolsep}{3.5pt} 
L_\mathcal{G} &=
\left[
\begin{array}{rrrrrrrr}
3  &  -1 &   -1  &   0  &   0  &   0  &   0 &   -1\\
-1  &   4  &  -1  &  -1  &  -1  &   0   &  0   &  0\\
-1  &  -1  &   3   & -1 &    0 &    0   &  0  &   0\\
0   & -1   & -1 &    3 &   -1  &   0 &    0  &   0\\
0  &  -1 &    0  &  -1  &   4  &  -1 &   -1   &  0\\
0  &   0  &   0  &   0  &  -1  &   3  &  -1  &  -1\\
0  &   0  &   0  &   0  &  -1  &  -1  &   3  &  -1\\
-1  &   0  &   0   &  0  &   0  &  -1  &  -1  &   3 
\end{array}\right]
\end{align}
and $D =\text{diag}(1, 1, 0, 0, 1, 0, 0, 2 )$.
With such $L_{\mathcal G}$ and $D$, we select the state-feedback gains $K_1$ for leaderless consensus and $K_2$ for leader-follower consensus, which can be found above.

For leaderless consensus, since $\rho(J(1))=0.77 $, by Theorem \ref{Theorem 1}, we choose $d_0=0.785$, $C_2=1.7977$, and $\gamma_1=0.8$ and $\gamma_2=6.7244$. With such parameters, the number of quantization levels should satisfy $2R + 1 \ge  10222$, which can be encoded by 14 bits, and the sufficient DoS-bound condition for consensus is $1/T + \Delta / \tau_D < 0.1048$. 
For leader-follower consensus, since $\rho(\widetilde P(1)) =0.9485$, according to Theorem 2, we choose $\tilde d_0=0.96$, $\tilde C_4=2.2247$, and $\gamma_1=0.965$ and $\gamma_2=7.96$. The number of quantization levels must satisfy $2R + 1 \ge 15150$ and can be encoded by 14 bits. The theoretical DoS-bound sufficient condition for leader-follower consensus is $1/T + \Delta / \tau_D < 0.0169$.

\begin{figure}[t]
	\begin{center}
		\includegraphics[width=0.44 \textwidth]{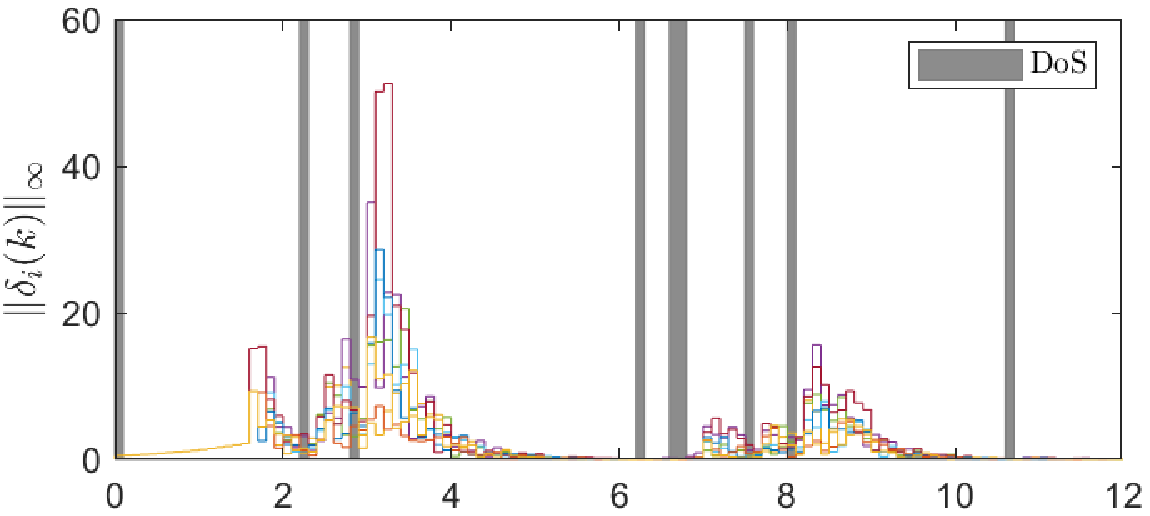}  \\
		\vspace{-0mm}
		\includegraphics[width=0.448 \textwidth]{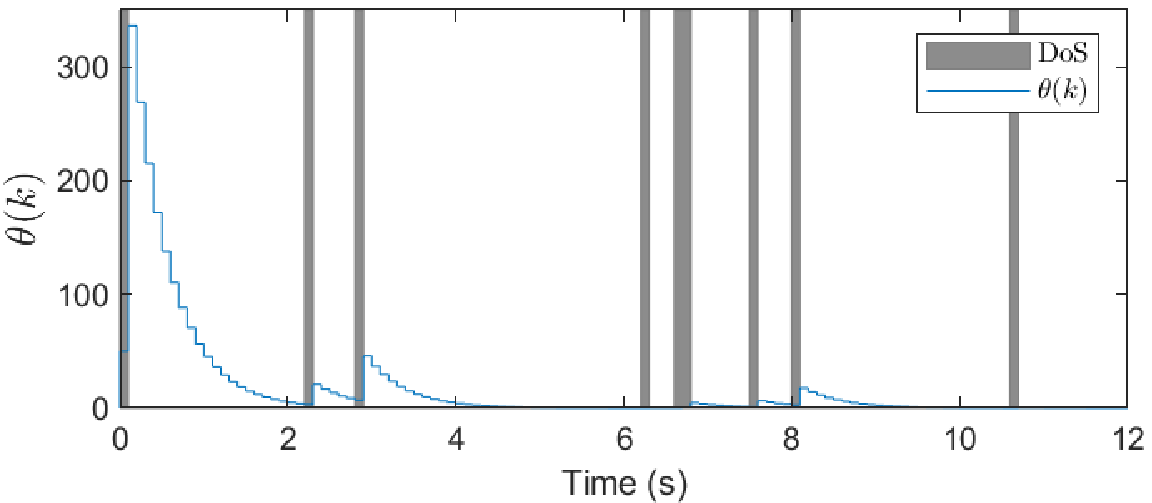} \\
		\vspace{-2mm}
		\linespread{1}\caption{Top: Time response of $\|\delta_i(k)\|_\infty$ in leaderless consensus; Bottom: Time response of $\theta(k)$ in leaderless consensus.} \label{leaderless}
	\end{center}
\end{figure}
\begin{figure}[t]
	\vspace{-3mm}
	\begin{center}
		\includegraphics[width=0.445 \textwidth]{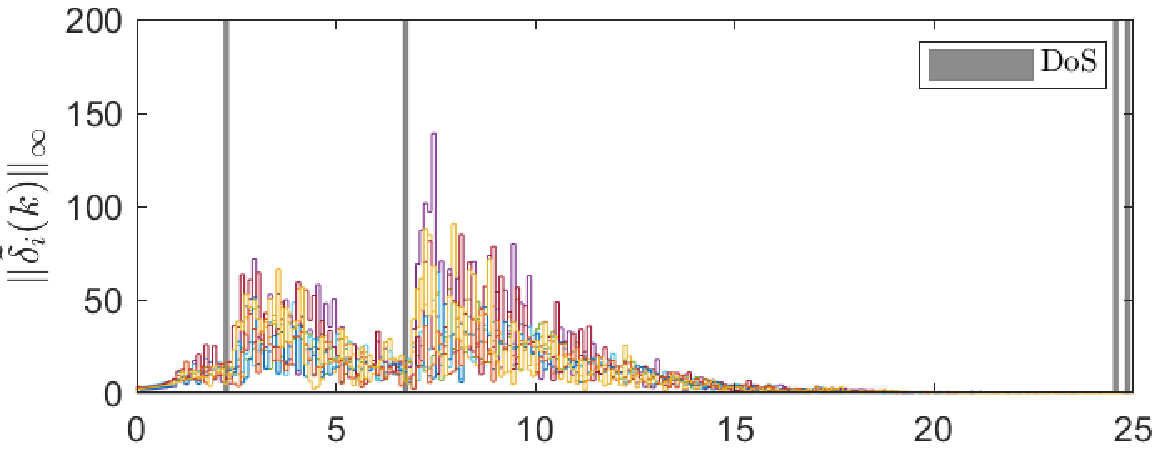}  \\
		\vspace{-3mm}
		\includegraphics[width=0.44 \textwidth]{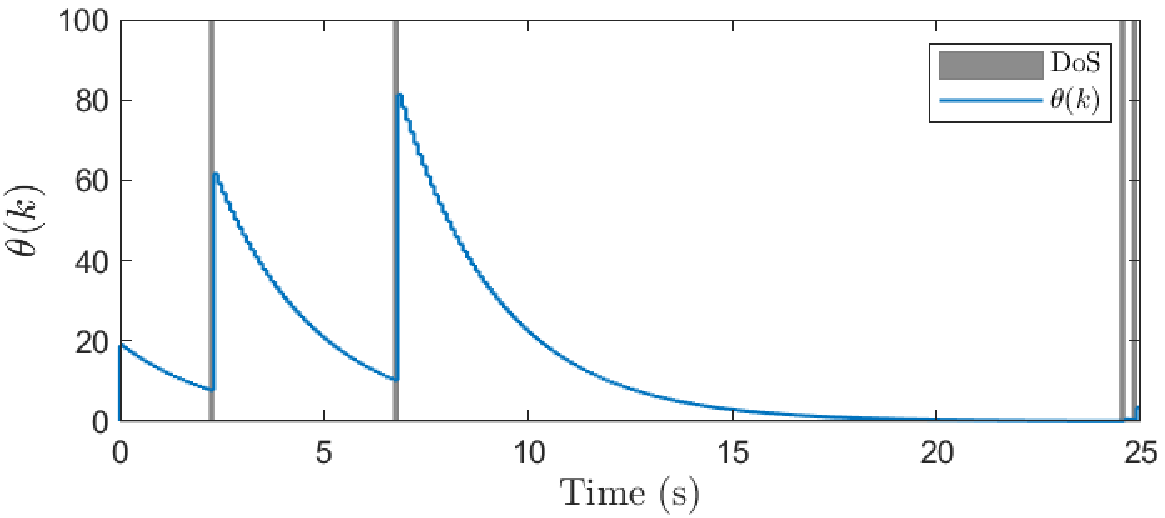} \\
		\vspace{-3mm}
		\linespread{1}\caption{Top: Time response of $\|\widetilde \delta_i(k)\|_\infty$ in leader-follower consensus; Bottom: Time response of $\theta(k)$ in leader-follower consensus. } \label{leaderfollower}
	\end{center}
	\vspace{-6mm}
\end{figure}

The time responses of $\|\delta_i(k)\|_\infty$ and $\theta(k)$ for leaderless consensus, and those of $\|\widetilde \delta_i(k)\|_\infty$ and $\theta(k)$ for leader-follower consensus are presented in Figs. \ref{leaderless} and \ref{leaderfollower}, respectively, in which the DoS attacks are generated randomly.
In Fig. \ref{leaderless} over the time horizon $12$s, the DoS 
signal yields $|\Xi(0,12)|=0.9$s and $n(0,12)=8$. 
This corresponds to averaged values
of $\tau_D\approx 1.5$ and $T\approx 13.33$, and $1/T + \Delta / \tau_D \approx 0.1417 $ for the case of leaderless consensus. Similarly, in Fig. \ref{leaderfollower}, the DoS 
signal yields $|\Xi(0,25)|=0.4$s and $n(0,12)=4$. 
This corresponds to averaged values
of $\tau_D\approx 6.25$ and $T\approx 62.5$, and $1/T + \Delta / \tau_D \approx  0.032 $ for the case of leader-follower consensus. Though the theoretical bounds regarding $1/T + \Delta/\tau_D$ are violated, by the first plots in Figs. \ref{leaderless} and \ref{leaderfollower}, respectively, one can see that both $\|\delta_i(k)\|_\infty$ and $\|\widetilde \delta_i(k)\|_\infty$ converge to zero. This implies that both the leaderless and leader-follower consensus are still successfully achieved.

The developed dynamic quantization with zooming-in and out capabilities can be clearly seen from the second plots in Figs. \ref{leaderless} and \ref{leaderfollower}. One can see that $\theta(k)$ increases when transmissions fail due to the presence of DoS, and decreases during the DoS-free periods.  
Meanwhile in the leaderless consensus simulation, the actual quantization output (i.e., $Q_R(\cdot)$) ranges from $-6$ to $6$ during the simulation. This amounts to the number of quantization levels 13, which is much smaller than the corresponding theoretical value 10222.
In the leader-follower consensus simulation, the actual followers' quantizer output ranges only from $-8$ to $6$ ($15$ quantization levels), and the quantization for the leader state takes only the values $-1, 0$ and $1$ ($3$ quantization levels). This is also much smaller than the obtained theoretical value 15150.


\section{Conclusions}
In this paper, we have presented results for the leaderless and leader-follower consensus problems of linear multi-agent systems with general dynamics under network data rate limitation and malicious DoS attacks. The design of quantized controller and the characterization of DoS attacks for consensus have been given. In particular, we have provided a feasible way of designing dynamic quantized control with zooming-in and zooming-out capabilities for the multi-agent systems with general dynamics, and such dynamic quantization makes finite data rate control possible without quantizer overflow under malicious DoS attacks. We have then characterized the bound of DoS attacks under which consensus of the multi-agent systems can be guaranteed. Discussions have been given on the trade-offs between bit rates and robustness against DoS. 

The results in this paper can be extended in several directions. 
One can consider to relax the assumption on the global knowledge about the communication topology by referring to \cite{li2014designing, xu2019distributed} and also consider the case of digraph by referring to \cite{feng2017distributed}.
It is also worthwhile considering the case of transmission delays\cite{xu2018consensusability}. 

\appendix

\subsection{Proof of Lemma \ref{Lemma3}}
In view of the dynamics of $\alpha$ in (\ref{alpha DoS}) and $\xi$ in (\ref{xi DoS}), it is easy to obtain such a form 
\begin{align}\label{B2}
\left[\begin{array}{ll}
\alpha(k+m) \\
\xi (k+m)
\end{array}
\right]
= \frac{\overline A(m)}{\gamma_2 ^ m} 
\left[\begin{array}{ll}
\alpha(k) \\
\xi (k)
\end{array}
\right]
\end{align}
where $ 0 \le m \le M$ (in Lemma \ref{Lemma Q}) denotes the number of consecutive unsuccessful transmissions after $k$ and $\overline A(m)$ is given in (\ref{bar A}). 
If $k+m+1$ is an instant of successful transmission, in view of (\ref{alpha no DoS}) and (\ref{B2}), one has
\begin{align}\setlength{\arraycolsep}{1pt}  \label{alpha k+m+1}
\alpha(k+m+1) &= (\left[\,G \,\,\,L\,\right] / \gamma_1)
[
\alpha^T(k+m) \quad
\xi ^ T(k+m)
]^T
\nonumber\\
& =  (\left[\,G \,\,\,L\,\right]  / \gamma_1)
(\overline A(m) / \gamma_2 ^ m  )
[ 
\alpha^T (k)\quad
\xi^T  (k)
]^T  \nonumber\\
&= \frac{G(m+1)}{\gamma_1} \alpha(k) + \frac{L(m+1)}{\gamma_1} \xi(k)
\end{align}  
with $G(m+1) $ and $L(m+1)$ in (\ref{G(m+1)}) and (\ref{L(m+1)}), respectively.

It is worth mentioning that (\ref{alpha k+m+1}) is a general form to incorporate the scenarios of successful and unsuccessful transmissions. 
If $m=0$, then in view of (\ref{bar A}), $\overline A_{11}(m)$ and $\overline A_{22}(m)$ become identity matrices and $\overline A_{12}(m)$ and $\overline A_{21}(m)$ are matrices with all zero entries. That is, $m=0$ indicates zero unsuccessful transmission between $k$ and $k+1$, and hence (\ref{alpha k+m+1}) is reduced to (\ref{alpha no DoS}) as a nominal update situation. 

Recall the unitary matrix $U$ in (\ref{Phi}), where one has 
$
U^T L_{\mathcal{G}} U = \text{diag}(0, \lambda_2, \cdots, \lambda_N)
$.
It is easy to verify that
$
(U \otimes I_n)^T (I_N \otimes A - L_{\mathcal{G}} \otimes BK)(U \otimes I_n)
=\, \text{diag}(A, A-\lambda_2 BK, \cdots, A- \lambda_N BK)
$.
With such $U$, we let
$
\overline \alpha(k) :=  (U \otimes I_n)^T  \alpha(k) = \left[ \overline \alpha_1 ^T (k) \,\, \overline \alpha_2 ^T (k)\right] ^T $
and let $\overline \xi(k, m+1)$ depending on $k$ and $m+1$ be
$
\overline \xi(k,m+1):= (U \otimes I_n)^T  L(m+1) \xi (k)  
=   \left[ \overline \xi_1 ^T (k,m+1) \,\, \overline \xi_2 ^T (k,m+1)\right] ^T  
$,
where $\overline \alpha_1 (k)$ and $\overline \xi_1(k,m+1)$ represent vectors with the first $n$ elements of $\overline \alpha(k)$ and $\overline \xi(k,m+1)$, respectively. One can verify that $\overline \alpha_1(k) = \mathbf{0}$ for all $k$. Equation
(\ref{alpha k+m+1}) can be transformed to
\begin{align}\label{bar alpha k+m+1}
\overline \alpha(k+m+1)
&= \frac{\overline G(m+1)}{\gamma_1} \overline \alpha(k) + \frac{ (U \otimes I_n)^T  L(m+1) }{\gamma_1} \xi (k) \nonumber\\
&= \frac{\overline G(m+1)}{\gamma_1} \overline \alpha(k) +\frac{1}{\gamma_1} \overline \xi(k,m+1)
\end{align} 
where $\overline G (m+1)$ is given in (\ref{bar G}). Note that $\overline G (m+1)$ is a block diagonal matrix. If we let $\bar \nu(k)= (U \otimes I _n) ^ T \xi (k)$, then (\ref{alpha DoS}) and (\ref{xi DoS}) can be rewritten as  
	\begin{align}
	\bar \alpha(k) &= \frac{\bar D}{\gamma_2} + (U \otimes I _n) ^T \frac{L}{\gamma_2}  (U \otimes I _n) \bar \nu (k-1) \label{72} \\
	\bar \nu(k) &= (U \otimes I _n) ^T \frac{H}{\gamma_2}  (U \otimes I _n) \bar \nu(k-1)  \nonumber\\
	& \quad - (U \otimes I _n) ^T \frac{L}{\gamma_2}  (U \otimes I _n) \bar \alpha(k-1)  \label{73}
	\end{align}
	in which $\bar D= \text{diag}(A, A-\lambda_2 BK, \cdots, A-\lambda_N BK)$. Similarly, one can obtain the equations corresponding to (\ref{alpha no DoS}) and (\ref{xi no DoS}) in terms of $\bar \alpha$ and $\bar \nu$. The analysis from (\ref{B2}) to (\ref{bar alpha k+m+1}) presents the compact calculation of the iteration between (\ref{72})--(\ref{73}), and those corresponding to (\ref{alpha no DoS})--(\ref{xi no DoS}). One can verify that $(U \otimes I _n) ^T L   (U \otimes I _n) $ and $(U \otimes I _n) ^T H  (U \otimes I _n) $ and $\bar D$ are block diagonal matrices, which implies $\overline G (m+1)$ block diagonal.

Recall that matrix $J(m+1) \in  \mathbb{R} ^{n(N-1) \times n(N-1) }$ (in (\ref{J})) denotes the remaining parts of $\overline G(m+1)$ after deleting the first $n$ rows and columns from $\overline G(m+1)$. Then one can obtain the following equation from (\ref{bar alpha k+m+1}) such that
$
\overline \alpha_2(k+m+1) = \frac{J(m+1)}{\gamma_1} \overline \alpha_2(k) +\frac{1}{\gamma_1} \overline \xi_2(k,m+1).
$
Recall that $s_r$ denotes the instant of successful transmissions for $r=0, 1, \cdots$, and $s_{-1}$ denotes $k=0$. Thus we have $s_{r} = k+m+1$, and $s_{r-1}= k$ if $k$ is a successful transmission instant.
Hence one has
\begin{align}\label{baralpha2}
\overline \alpha_2(s_{r}) = \frac{J(m+1)}{\gamma_1} \overline \alpha_2(s_{r-1}) +\frac{1}{\gamma_1} \overline \xi_2(s_{r-1},m+1).
\end{align}  
For distinguishing $J(m+1)$ in iteration steps, we let $ J_{r-1}(m_{r-1} + 1)$ denote the $J(m+1)$ in (\ref{baralpha2}) used for the iteration from $s_{r-1}$ to $s_r$ with $s_r - s_{r-1} = (m_{r-1}+1) \Delta$. To reduce notation burden, we further let $J_{r-1}$ represent $ J_{r-1}(m_{r-1} + 1)$. Then (\ref{baralpha2}) is written as
$
\overline \alpha_2(s_{r}) = \frac{J_{r-1}}{\gamma_1} \overline \alpha_2(s_{r-1}) +\frac{1}{\gamma_1} \overline \xi_2(s_{r-1},m+1) \nonumber
$
for $r=0, 1, \cdots$.
By iteration, it is easy to obtain 
\begin{align}\label{B9}
\overline \alpha_2(s_{r})
=& \prod_{p=0}^{r} \frac{J_{p-1}}{\gamma_1} \overline \alpha_2(s_{-1}) + \sum_{p=0}^{r-1}\left( \prod_{q=p}^{r-1}\frac{J_q}{\gamma_1}  \right) \frac{\overline \xi_2(s_{p-1},m+1)}{\gamma_1}\nonumber\\
&+ \frac{\overline \xi_2 (s_{r-1}, m+1)}{\gamma_1}.
\end{align}
In case the networked multi-agent systems are not subject to DoS attacks, then $J_{p-1}$ and $J_q$ in (\ref{B9}) are equal to $J(1)$, and there exist $C_2   \ge 1$ and $ \rho (J(1)) < d_0 <1$ such that $\|(J(1))^p\| \le C_2 d_0 ^ p $ ($p=1, 2, \cdots$). This implies that $\|(J(1)/\gamma_1)^p\| \le C_2 d^p$, where $\gamma_1 > d_0$ and $0<d = d_0 /\gamma_1 <1$. Therefore, the type of calculation reduces to the one in \cite{you2011network}.

Recall that we have selected $\gamma_2$ in Lemma \ref{Lemma3} which can make $\| J(m+1) \| \le  \rho(J(1))/C_2$ hold for $m=1, \cdots, M$. By such $\gamma_2$, one has that the iteration of $J_{p-1}/\gamma_1$ in (\ref{B9}) yields
\begin{align}\label{B10}
\|\prod_{p=0}^{r}  (J_{p-1}/\gamma_1)  \| \le   \prod_{p=0}^{r}   \|(J_{p-1}/\gamma_1)  \|  \le   C_2 d^ r.
\end{align}
Notice that $C_2$ does not accumulate in the iteration because the $C_2$ caused by $\|(J(1)/\gamma_1)^p\| \le C_2 d^p$ (iteration in a DoS-free interval) is canceled out by the $C_2$ in $\| J(m+1)/\gamma_1\| \le  \rho(J(1))/( \gamma_1 C_2 )  < d_0 / (\gamma_1 C_2) = d/C_2 $ ($m = 1, \cdots, M$ representing the number of iteration during a DoS interval).

By Lemma \ref{Lemma3}, we have selected $d_0 < \gamma_1 < 1$ and 
$
\theta_0 \ge C_{x_0} \gamma_1 / \sigma
$.
By such $\theta_0$, we have
$
\| \alpha(0)    \|  = \| \delta(0)\|    / \theta_0 
\le \sqrt{Nn} \|\delta(0)\|_\infty / \theta_0  
\le 2 \sqrt{Nn}  C_{x_0}/ \theta_0 \le 2\sqrt{Nn}   \sigma  / \gamma_1 
$
where we use the fact $\|\delta(0)\|_\infty \le 2 C_{x_0} $. By noting that $\|(U \otimes I_n )^T \| =1$, we have that $\|\overline \alpha_2(s_{-1}) \|$ satisfies
\begin{align}\label{bar alpha2(0)}
\|\overline \alpha_2 (s_{-1}) \| &= \|\overline \alpha_2 (0) \| \le \| \overline \alpha(0) \| \nonumber\\
&\le \| (U \otimes I_n )^T \| \|\alpha(0) \| 
\le   2\sqrt{Nn}  \sigma  / \gamma_1.  
\end{align}
Furthermore, one has
$
\|\xi(s_{-1})\|_\infty 
= 
\|\xi(0)\|_\infty 
\le \left\|    (\hat x(0) -x(0))/\theta_0   \right\|_\infty 
= \left\|    x(0)/\theta_0    \right\|_\infty \le C_{x_0} / \theta_0 \le \sigma / \gamma_1.
$
By assumption, we have $\|\xi (s_p)\|_\infty \le \sigma/\gamma_1 $ for $p=0, 1, \cdots, r$. Incorporating $\|\xi(s_{-1})\|_\infty$, overall one has $\|\xi (s_p)\|_\infty \le \sigma/\gamma_1 $ for $p=-1, 0, \cdots, r$. Hence, we obtain 
\begin{flalign}\label{B13}
\|\overline \xi_2(s_p, m+1)\| &\le \|(U \otimes I_n)^T L (m+1)\|\|\xi(s_{p})\| \nonumber\\
& = \|L (m+1)\|\|\xi(s_{p})\| \le C_0 \sqrt{N n} \sigma /\gamma_1    &&
\end{flalign}
for $p=-1, 0, \cdots$, where $C_0$ is given by (\ref{C5}).

Substituting (\ref{B10}), (\ref{bar alpha2(0)}) and (\ref{B13}) into (\ref{B9}), we have
$
\|\overline \alpha_2 (s_r)  \| 
\le 2 C_2  \sqrt{Nn} \frac{\sigma}{\gamma_1} d^{r }   + \frac{C_0 C_2 \sqrt{Nn} \sigma}{ (1-d) \gamma_1 ^ 2} (1 -   d^r) 
\le C_1  \sigma/ \gamma_1
$ for $r= 0, 1, \cdots$, where $C_1$ is as in (\ref{C6}).
Incorporating (\ref{bar alpha2(0)}), it is obvious that 
$
\|\alpha(s_r)\| \le \|((U \otimes I_n) ^T) ^ {-1}\| \|\overline \alpha(s_r)\|  
=\|\overline \alpha_2(s_r) \|  \le C_1 \sigma /\gamma_1, \quad r= -1, 0, \cdots
$
with the facts that $ \|((U  \otimes I_n) ^T) ^ {-1} \|= 1$ and $\overline \alpha_1   (k) = \mathbf{0}$. Finally, one has
\begin{align}
\| [\alpha^T(s_r) \,\,\,  \xi^T(s_r)  ]^T\|  
=& \sqrt{\|\alpha(s_r)\|^2 + \|\xi(s_r)\|^2} \nonumber\\
\le& \sigma \sqrt{C_1 ^2 +  Nn}  / \gamma_1 , \,\, r=-1, 0, \cdots \nonumber
\end{align}
where $\|\xi(s_r)\| \le \sqrt{Nn} \|\xi(s_r)\|_\infty \le \sqrt{Nn} \sigma/\gamma_1$. 
\qedp

\subsection{Proof of Lemma \ref{Lemma4}}
In view of (\ref{beta})--(\ref{epsilon 0}), one obtains the vector form as
\begin{align}\setlength{\arraycolsep}{1.5pt} \label{lf compact DoS}
\left[\begin{array}{c}
\beta(k)\\
\epsilon(k)\\
\epsilon_0(k) 
\end{array}\right] 
= \frac{\widetilde A }{\gamma_2} 
\left[\begin{array}{c}
\beta(k-1)\\
\epsilon(k-1)\\
\epsilon_0(k-1) 
\end{array}\right] 
\end{align}
where the matrix $\widetilde A$ is given in (\ref{tilde A}).
By the iterations of (\ref{lf compact DoS}), one has 
\begin{align}\label{5}
\left[\begin{array}{c}
\beta(k+m)\\
\epsilon(k+m)\\
\epsilon_0(k+m) 
\end{array}\right] 
= \frac{\widetilde A(m) }{\gamma_2 ^m} 
\left[\begin{array}{c}
\beta(k)\\
\epsilon(k)\\
\epsilon_0(k) 
\end{array}\right]
\end{align}
with $m=0, 1, \cdots, M$. The matrix $\widetilde A(m)$ is given in (\ref{tilde Am}).
If a successful transmission occurs at $k+m+1$, according to (\ref{beta no DoS}) and (\ref{5}), one has
\begin{align}\label{beta DoS}
&\, \beta(k+m + 1) \nonumber\\
&=\, \frac{[\Pi \quad \Sigma \,\, -\Phi]}{\gamma_1}
\left[\begin{array}{ccc}
\beta ^T(k+m)&
\epsilon ^T(k+m)&
\epsilon_0 ^T(k+m) 
\end{array}\right]^T \nonumber\\
&=\, \frac{[\Pi \quad \Sigma \,\, -\Phi]}{\gamma_1} 
\frac{\widetilde A(m) }{\gamma_2 ^m} 
\left[\begin{array}{ccc}
\beta^T(k)&
\epsilon^T(k)&
\epsilon_0 ^T (k) 
\end{array}\right]^T  \nonumber\\
&=\, \frac{P(m+1)}{\gamma_1} \beta(k) + \frac{S(m+1)}{\gamma_1} \epsilon(k) - \frac{Z(m+1)}{\gamma_1}  \epsilon_0(k)
\end{align}
where the matrices
$
P(m+1) =  (\Pi \widetilde A_{11} (m) + \Sigma \widetilde A_{21}(m))/\gamma_2 ^m$, 
$S(m+1)  =  ( \Pi \widetilde A_{12} (m) + \Sigma \widetilde A_{22}(m) )/ \gamma_2 ^m$
and 
$Z(m+1) = ( \Pi\widetilde A_{13}(m) + \Sigma \widetilde A_{23}(m)  - \Phi (I_N \otimes A^m) ) / \gamma_2 ^m$ are given after (\ref{tilde Am}).

There exists a unitary matrix $\widetilde \Psi$ such that $\widetilde \Psi ^ {-1} (L _{\mathcal G} + D) \widetilde \Psi$ is an upper-triangular matrix whose diagonals are the eigenvalues of the ones of $L _{\mathcal G} + D$. With such $\tilde \Phi$, we obtain the matrices $\widetilde P(m+1)$ in (\ref{tilde P}) and 
\begin{align}
\widetilde S(m+1) & = (\widetilde \Psi \otimes I_n)^T S(m+1) \\
\widetilde Z(m+1) & = (\widetilde \Psi \otimes I_n)^T Z(m+1).
\end{align}
In case $m=0$, $\widetilde A_{11} (0)$ becomes the identity matrix and $\widetilde A_{21}(0)$ is a matrix with all zero entries. Then $\widetilde P(m+1)$ is reduced to
$
\widetilde P (1) = (\widetilde \Psi \otimes I_n)^T P(1)(\widetilde \Psi \otimes I_n) 
=  (\widetilde \Psi \otimes I_n)^T (\Pi \widetilde A_{11} (0) + \Sigma \widetilde A_{21}(0)) (\widetilde \Psi \otimes I_n) 
=  (\widetilde \Psi \otimes I_n)^T   \Pi   (\widetilde \Psi \otimes I_n) 
$
as in (\ref{tilde P1}). 
By using the transformations
$
\widetilde \beta(k)  = (\widetilde \Psi \otimes I_n) ^ T \beta (k),  
\widetilde \epsilon(k, m+1 )  = \widetilde S (m+1) \epsilon (k)$ and 
$\widetilde \epsilon_0(k,m+1 )  = \widetilde Z (m+1) \epsilon_0 (k)$,
the equation (\ref{beta DoS}) can be rewritten as 
\begin{align}\label{tilde beta 0}
\widetilde \beta(k+m+1) &= \frac{\widetilde P(m+1)}{\gamma_1} \widetilde \beta (k) \nonumber\\
&\quad+ \frac{\widetilde \epsilon(k, m+1)}{\gamma_1} - \frac{\widetilde \epsilon_0(k, m+1)}{\gamma_1}. 
\end{align}
Let $\widetilde P$ denote any matrix in the set of matrices $\mathcal P$ as 
\begin{align}
\widetilde P \in \mathcal P = \{ \widetilde P(1), \cdots,\widetilde P(m+1), \cdots, \widetilde P(M+1)\}.
\end{align}
When $k$ and $k+m+1$ are instants of successful transmissions, by substituting $k$ and $k+m+1$ with $s_{r-1}$ and $s_r$, respectively, (\ref{tilde beta 0}) can be written as 
\begin{align}\label{tilde beta}
\widetilde \beta(s_r) = \frac{\widetilde P}{\gamma_1} \widetilde \beta (s_{r-1}) + \left(\frac{\widetilde \epsilon(s_{r-1},m+1)}{\gamma_1} - \frac{\widetilde \epsilon_0(s_{r-1},m+1)}{\gamma_1}  \right).
\end{align}

With (\ref{tilde beta}), we conduct the following analysis to obtain an upper bound of $\|\widetilde \beta (s_r)\|$.
First we compute
$
\|\widetilde \beta (s_{-1})\| \le   2\sqrt{Nn} \sigma / \gamma_1.
$
Then we can derive 
\begin{align}
&\|\widetilde \epsilon(s_{r-1},m+1)    - \widetilde \epsilon_0 (s_{r-1},m+1)  \| \nonumber\\
&=   \|\widetilde S(m+1) \epsilon(s_{r-1})   - \widetilde Z (m+1)   \epsilon_0(s_{r-1}) \| \nonumber\\
&= \| (\widetilde \Psi \otimes I_n)^T S(m+1) \epsilon(s_{r-1})  \nonumber\\
&\,\,\,\, -(\widetilde \Psi \otimes I_n)^T Z(m+1)  \epsilon_0(s_{r-1})		\| \nonumber\\
&\le  \|S(m+1)\| \|\epsilon(s_{r-1})\|  +  \|Z(m+1)\| \| \epsilon_0(s_{r-1})\|  \nonumber\\
&=  \sqrt{Nn} \widetilde C_0 \sigma/ \gamma_1 + \sqrt{Nn} \widetilde C_1 \sigma/\gamma_1 = \sqrt{Nn} \widetilde C_2 \sigma/ \gamma_1 
\end{align}
where $\widetilde C_2 = \widetilde C_0 + \widetilde C_1$ with 
$
\widetilde C_0 = \max_{m=0, 1, \cdots, M} \|S(m+1)\|$ and $\widetilde C_1 = \max_{m=0, 1, \cdots, M} \|Z(m+1)\|.
$

Now we analyze the iteration of (\ref{tilde beta}). First, there exist $\widetilde C_4 \ge 1$ and $\rho(\widetilde P(1))<\widetilde d_0<1$ such that $\|\widetilde P (1) ^ p\| \le \widetilde C_4 \widetilde d_0 ^ p$ for $p=1, 2, \cdots$. This implies that $\|(\widetilde P (1)    / \gamma_1 )^ p\| \le \widetilde C_4 (\widetilde d_0 /\gamma_1) ^ p = \widetilde C_4 \widetilde d ^ p$ where $\widetilde d_0<\gamma_1<1$ and $\widetilde d = \widetilde d_0 / \gamma_1$. Recall that we have selected $\gamma_2$ in Lemma \ref{Lemma4} such that the induced $\| \widetilde P(m+1)  \| \le  \rho(\widetilde P(1))/ \widetilde C_4$, in which $m=1, \cdots, M$ representing the presence of DoS. This implies that $\| \widetilde P(m+1) / \gamma_1 \| \le  \rho(\widetilde P(1))/ (\gamma_1 \widetilde C_4) <  \widetilde d_0 / (\gamma_1 \widetilde C_4) <    \widetilde d  / \widetilde C_4$ for $m=1, \cdots, M$.
Then, combining the cases of $\widetilde P(1)$ and $\widetilde P (m+1)$ for $m=1,\cdots, M$, one has that 
\begin{align}\label{96}
\left\|    \left( \widetilde P /\gamma_1 \right) ^ k \right\|  \le  \widetilde C _4  \widetilde d ^k,
\end{align}
where $\widetilde C_4$ induced by $\|\widetilde P (1) ^ p\| \le \widetilde C_4 \widetilde d_0 ^ p$ is not accumulated in the iteration (\ref{96}) since the $\widetilde C_4$ is canceled out by the $\widetilde C_4$ in $\| \widetilde P(m+1) / \gamma_1 \| <  \widetilde d  / \widetilde C_4$.
In view of (\ref{tilde beta}) and following the very similar calculation as in the proof of Lemma \ref{Lemma3}, one has
$
\|\widetilde \beta (s_r)\| \le \widetilde C_3 \sigma / \gamma_1
$ with $\widetilde C_3$ in (\ref{C9}) and furthermore
$
\|\beta (s_r)\| \le  \| (( \widetilde \Psi  \otimes I_n)^T )^{-1}  \| \|\widetilde \beta (s_r)\| \le \widetilde C_3 \sigma/ \gamma_1
$.
Moreover, we also have
$
\|\epsilon(s_r) \| \le \sqrt{Nn} \| \epsilon (s_r)\| _\infty \le \sqrt{Nn} \sigma / \gamma_1
$ and similarly $\|\epsilon_0 (s_r)\| \le \sqrt{Nn}\sigma /\gamma_1$.
Eventually, one has
\begin{align}
\|[\beta ^T (s_r)\,\,\, \epsilon^T(s_r)\,\,\,   \epsilon_0 ^ T (s_r)  ]\|^T 
& \le   \sqrt{  \widetilde C_3 ^2 \sigma^2 / \gamma_1^2   + 2Nn \sigma^2 / \gamma_1^2 } \nonumber
\end{align}
for $r=-1, 0, \cdots$, and obtains the desired result (\ref{vector Lemma4}). \qedp

\bibliographystyle{IEEEtran}

\bibliography{multiquantization}

\end{document}